\documentclass[fleqn,usenatbib]{mnras}

\usepackage[T1]{fontenc}
\usepackage{ae,aecompl}

\usepackage{graphicx}	
\usepackage{amsmath}	
\usepackage{amssymb}	

\usepackage{epsfig}
\usepackage{epstopdf} 
\usepackage{fancyhdr}
\usepackage{multirow} 
\usepackage{multicol}
\usepackage{rotating}
\usepackage{url,times,amsfonts,color}
\usepackage{txfonts,float}
\usepackage{titlesec}
\titlespacing*{\section}{0pt}{5mm}{1mm}
\titlespacing*{\subsection}{0pt}{3mm}{1mm}
\titlespacing*{\subsubsection}{0pt}{3mm}{1mm}

\usepackage{xspace}

\newcommand{\galacticus}{\textsc{Galacticus}\xspace}

\newcommand{\MD}{\textsc{MultiDark}\xspace}
\newcommand{\bmd}{\textsc{BigMDPL}\xspace}
\newcommand{\blc}{\textsc{BigMD-LC}\xspace}
\newcommand{\mlc}{\textsc{MD-LC}\xspace}
\newcommand{\MDG}{\textsc{MultiDark-Galaxies}\xspace}

\newcommand{\MDPL}{\textsc{MDPL2}\xspace}
\newcommand{\MDgal}{\textsc{MDPL2}-\texttt{Galacticus}\xspace}

\newcommand{\consistenttree}{\textsc{Consistent Trees}}

\newcommand{\rockstar}{\textsc{Rockstar}}

\newcommand{\sdss}{\texttt{SDSS}\xspace}
\newcommand{\boss}{\texttt{BOSS}\xspace}
\newcommand{\bossDR}{\texttt{BOSS-CMASS DR12}\xspace}
\newcommand{\cmass}{\texttt{CMASS}\xspace}
\newcommand{\cmDR}{\texttt{CMASS DR12}\xspace}

\newcommand{\all}{\texttt{Gal-all}\xspace}
\newcommand{\cm}{\texttt{Gal-cols}\xspace}
\newcommand{\den}{\texttt{Gal-dens}}
\newcommand{\ma}{\texttt{Gal-mass}\xspace}

\newcommand{\allS}{\texttt{Gal-all} sample\xspace}
\newcommand{\cmS}{\texttt{Gal-cols} sample\xspace}
\newcommand{\denS}{\texttt{Gal-dens} sample\xspace}
\newcommand{\maS}{\texttt{Gal-mass} sample\xspace}

\newcommand{\SMF}{\textsc{\textit{SMF}}\xspace}
\newcommand{\SMFs}{\textsc{\textit{SMFs}}\xspace}

\newcommand{\SFR}{\textsc{\textit{SFR}}\xspace}
\newcommand{\SFRs}{\textsc{\textit{SFRs}}\xspace}

\newcommand{\sSFR}{\textsc{\textit{sSFR}}\xspace}

\newcommand{\SHMF}{\textsc{\textit{SHMF}}\xspace}
\newcommand{\SHMFs}{\textsc{\textit{SHMFs}}\xspace}

\newcommand{\twoPCF}{\textsc{\textit{2pCF}}\xspace}
\newcommand{\twoPCFs}{\textsc{\textit{2pCFs}}\xspace}

\newcommand{\ri}{\textit{(r-i)}\xspace}
\newcommand{\gr}{\textit{(g-i)}\xspace}
\newcommand{\gi}{\textit{(g-i)}\xspace}
\newcommand{\iband}{$i$-band\xspace}

\newcommand{\gband}{$g$-band\xspace}

\newcommand{\Mbnd}{{\ifmmode{M_{\rm bnd}}\else{$M_{\rm bnd}$}\fi}}
\newcommand{\Mfof}{{\ifmmode{M_{\rm fof}}\else{$M_{\rm fof}$}\fi}}
\newcommand{\Mmean}{{\ifmmode{M_{\rm 200m}}\else{$M_{\rm 200m}$}\fi}}
\newcommand{\MBN}{{\ifmmode{M_{\rm BN98}}\else{$M_{\rm BN98}$}\fi}}
\newcommand{\Rc}{{\ifmmode{R_{\rm 200c}}\else{$R_{\rm 200c}$}\fi}}
\newcommand{\Vz}{{\ifmmode{V_{\rm eff}}\else{$V_{\rm eff}$\xspace}\fi}}
\newcommand{\ltsima}{$\; \buildrel < \over \sim \;$}
\newcommand{\gtsima}{$\; \buildrel > \over \sim \;$}
\newcommand{\lsim}{\lower.5ex\hbox{\ltsima}}
\newcommand{\gsim}{\lower.5ex\hbox{\gtsima}}

\newcommand{\hGpc}{{\ifmmode{h^{-1}{\rm Gpc}}\else{$h^{-1}$Gpc\xspace}\fi}}
\newcommand{\hMpc}{{\ifmmode{h^{-1}{\rm Mpc}}\else{$h^{-1}$Mpc\xspace}\fi}}
\newcommand{\hkpc}{{\ifmmode{h^{-1}{\rm kpc}}\else{$h^{-1}$kpc\xspace}\fi}}
\newcommand{\hMsun}{{\ifmmode{h^{-1}{\rm {M_{\odot}}}}\else{$h^{-1}{\rm{M_{\odot}}}$}\fi}}

\newcommand{\Mstar}{{\ifmmode{M_{*}}\else{$M_{*}$\xspace}\fi}}
\newcommand{\Mhalo}{{\ifmmode{M_{\rm Halo}}\else{$M_{\rm Halo}$\xspace}\fi}}
\newcommand{\Mhot}{{\ifmmode{M_{\rm Hot}}\else{$M_{\rm Hot}$\xspace}\fi}}
\newcommand{\Mcold}{{\ifmmode{M_{\rm Cold}}\else{$M_{\rm Cold}$\xspace}\fi}}
\newcommand{\Mc}{{\ifmmode{M_{\rm 200c}}\else{$M_{\rm 200c}$\xspace}\fi}}
\newcommand{\Mvir}{{\ifmmode{M_{\rm vir}}\else{$M_{\rm vir}$\xspace}\fi}}
\newcommand{\Mbh}{{\ifmmode{M_{\rm BH}}\else{$M_{\rm BH}$\xspace}\fi}}
\newcommand{\Zcold}{{\ifmmode{Z_{\rm Cold}}\else{$Z_{\rm Cold}$\xspace}\fi}}

\newcommand{\Ngal}{{\ifmmode{N_{\rm gal}}\else{$N_{\rm gal}$}\fi}}
\newcommand{\Norph}{{\ifmmode{N_{\rm orphan}}\else{$N_{\rm orphan}$}\fi}}
\newcommand{\Nxorph}{{\ifmmode{N_{\rm non-orphan}}\else{$N_{\rm non-orphan}$}\fi}}
\newcommand{\Zsolar}{{\ifmmode{Z_{\odot}}\else{$Z_{\odot}$}\fi}}
\newcommand{\Msun}{{\ifmmode{{\rm {M_{\odot}}}}\else{${\rm{M_{\odot}}}$\xspace}\fi}}
\newcommand{\Msunyr}{{\ifmmode{{\rm {M_{\odot}yr^{-1}}}}\else{${\rm{M_{\odot}yr^{-1}}}$\xspace}\fi}}

\newcommand{\MpcCu}{{\ifmmode{{\rm Mpc}^3}\else{${{\rm Mpc}^3}$\xspace}\fi}}
\newcommand{\MpcV}{{\ifmmode{{\rm Mpc}^{-3}}\else{${{\rm Mpc}^{-3}}$\xspace}\fi}}
\newcommand{\logT}{{\ifmmode{\log_{\rm 10}}\else{$\log_{\rm 10}$\xspace}\fi}}
\newcommand{\dex}{{\ifmmode{\rm dex^{-1}}\else{$\rm dex^{-1}$\xspace}\fi}}

\newcommand{\dperp}{$d_{\perp}$\xspace}

\newcommand{\Mr}{{\ifmmode{{{M_{\rm r}}}}\else{${{M_{\rm r}}}$\xspace}\fi}}
\newcommand{\rp}{{\ifmmode{{{r_{\rm p}}}}\else{${{r_{\rm p}}}$\xspace}\fi}}
\newcommand{\pim}{{\ifmmode{{{\pi_{\rm max}}}}\else{${{\pi_{\rm max}}}$\xspace}\fi}}

\newcommand{\Mcut}{{\ifmmode{{{M_{\rm cut}}}}\else{${{M_{\rm cut}}}$\xspace}\fi}}
\newcommand{\Mmin}{{\ifmmode{{{M_{\rm min}}}}\else{${{M_{\rm min}}}$\xspace}\fi}}
\newcommand{\Mone}{{\ifmmode{{{M_{\rm 1}}}}\else{${{M_{\rm 1}}}$\xspace}\fi}}
\newcommand{\sigM}{{\ifmmode{{{\sigma_{\rm{log} \Mstar}}}}\else{${{\sigma_{\log_{\rm 10} \Mstar}}}$\xspace}\fi}}
\newcommand{\C}{{\ifmmode{{{C_{\rm NFW}}}}\else{${{C_{\rm NFW}}}$\xspace}\fi}}

\def\lesssim{\mathrel{\hbox{\rlap{\hbox{\lower4pt\hbox{$\sim$}}}\hbox{$<$}}}}
\def\gtrsim{\mathrel{\hbox{\rlap{\hbox{\lower4pt\hbox{$\sim$}}}\hbox{$>$}}}}

\newcommand{\Tab}[1]{Table~\ref{#1}}
\newcommand{\Sec}[1]{Section~\ref{#1}}
\newcommand{\App}[1]{Appendix~\ref{#1}}

\newcommand{\Eq}[1]{Eq.~(\ref{#1})}
\newcommand{\Eqs}[1]{Eqs.~(\ref{#1})}
\newcommand{\EqO}[1]{(\ref{#1})}
\newcommand{\Fig}[1]{Fig.~\ref{#1}}
\newcommand{\beq}{\begin{equation}}
\newcommand{\eeq}{\end{equation}}

\def\beqa{\begin{eqnarray}}
\def\eeqa{\end{eqnarray}}
\def\Mpc{${\rm Mpc}$}

\def\hMpc{$h^{-1}\,{\rm Mpc}$}
\def\hkpc{$h^{-1}\,{\rm kpc}$}

\title[SAM CMASS-mocks]
{A semi-analytical perspective on massive galaxies at $z\sim0.55$}
\author[Stoppacher et al. 2018]
        {D. Stoppacher,$^{1,2,\dagger}$\thanks{E-mail: doris.stoppacher@csic.es}
        F. Prada,$^{3}$
	    A. D. Montero-Dorta,$^{4}$			
        S. Rodr\'{\i}guez-Torres,$^{2}$ 	
\newauthor 
        A. Knebe,$^{2,5,6}$
        G. Favole,$^{7}$					
    	W. Cui,$^{2,8}$			       	
        A. J. Benson,$^{9}$   				
        C. Behrens,$^{10}$			  		
        A. A. Klypin$^{11}$                 
\\
\\
$^{1}$Instituto de F\'{\i}sica Te\'orica, (UAM/CSIC), Universidad Aut\'onoma de Madrid, Cantoblanco, E-28049 Madrid, Spain\\
$^{2}$Departamento de F\'isica Te\'{o}rica, M\'{o}dulo 15, Facultad de Ciencias, Universidad Aut\'{o}noma de Madrid, E-28049 Madrid, Spain\\
$^{3}$Instituto de Astrof\'{\i}sica de Andaluc\'{\i}a (CSIC), Glorieta de la Astronom\'{\i}a, E-18080 Granada, Spain\\
$^{4}$Departamento de F\'isica Matem\'atica, Instituto de F\'isica, Universidade de S\~ao Paulo, Rua do Mat\~ao 1371, CEP 05508-090, S\~ao Paulo, Brazil\\
$^{5}$Centro de Investigaci\'{o}n Avanzada en F\'isica Fundamental (CIAFF), Facultad de Ciencias, Universidad Aut\'{o}noma de Madrid, 28049 Madrid, Spain\\
$^{6}$International Centre for Radio Astronomy Research, University of Western Australia, 35 Stirling Highway, Crawley, Western Australia 6009, Australia\\
$^{7}$European Space Astronomy Centre (ESAC), 28692 Villanueva de la Ca\~nada, Madrid, Spain\\
$^{8}$Institute for Astronomy, University of Edinburgh, Royal Observatory, Edinburgh EH9 3HJ, United Kingdom\\
$^{9}$Institut f\"{u}r Astrophysik, Georg-August Universit\"at G\"{o}ttingen, Friedrich-Hund-Platz 1, 37077, G\"{o}ttingen, Germany\\
$^{10}$Carnegie Observatories, 813 Santa Barbara Street, Pasadena, CA 91101, USA\\
$^{11}$Astronomy Department, New Mexico State University, Dept.4500, Las Cruces, NM 88003-0001, USA\\
$^{\dagger}$Severo Ochoa IFT-CSIC Scholar
}
\date{Accepted XXX. Received YYY; in original form ZZZ}
\pubyear{2018}

\begin{document}
\label{firstpage}
\pagerange{\pageref{firstpage}--\pageref{lastpage}}
\maketitle


\begin{abstract}

\noindent The most massive and luminous galaxies in the Universe serve as powerful probes to study the formation of structure, the assembly of mass, and cosmology. However, their detailed formation and evolution is still barely understood. Here we extract a sample of massive mock galaxies from the semi-analytical model of galaxy formation (SAM) \galacticus from the \MDG, by replicating the \cmass photometric selection from the \sdss-\textsc{III} Baryon Oscillation Spectroscopic Survey (\boss). The comparison of the \galacticus CMASS-mock with \boss-\cmass data allows us to explore different aspects of the massive galaxy population at $0.5<z< 0.6$, including the galaxy-halo connection and the galaxy clustering. We find good agreement between our modelled galaxies and observations regarding the galaxy-halo connection, but our CMASS-mock over-estimates the clustering amplitude of the 2-point correlation function, due to a smaller number density compared to \boss, a lack of blue objects, and a small intrinsic scatter in stellar mass at fixed halo mass of $<0.1$ dex. To alleviate this problem, we construct an alternative mock catalogue mimicking the \cmass colour-magnitude distribution by randomly down-sampling the SAM catalogue. This CMASS-mock reproduces the clustering of \cmass galaxies within 1$\sigma$ and shows some environmental dependency of star formation properties that could be connected to the quenching of star formation and the assembly bias.

\end{abstract}
\noindent
\begin{keywords}
  methods: semi-analytical simulation -- galaxies: haloes -- galaxies:
  evolution -- cosmology: theory -- dark matter
\end{keywords}

\renewcommand{\thefootnote}{\textsuperscript{\arabic{footnote}}}

\section{Introduction} \label{sec:introduction}
The most luminous and massive galaxies in the Universe serve as powerful probes to study the formation of structure, the assembly of mass and cosmology, but their detailed formation and evolution, especially their connection to feedback processes, quenching of star formation or the assembly bias is still not sufficiently understood or quantified \citep[][]{Tinker13,Wechsler&Tinker18}. The Sloan Digital Sky Survey \texttt{SDSS-III}/Baryon Oscillation Spectroscopic Survey \citep[\boss,][]{Schlegel09_BOSS,Eisenstein11_SDSS3,Dawson13_BOSS} was dedicated to studying properties of the large-scale distribution of massive galaxies and provides a well studied sample of $\sim1.5$ million luminous red galaxies (LRGs). The \boss sample is divided into two: a low redshift (\texttt{LOWZ}) and a high redshift sample (\cmass, stands for ``constant mass''), respectively. \par

The \cmass sample covers a wide redshift range from $0.43<z<0.75$ exhibiting a peak in comoving number density of $n\sim3.4\times10^{-4}$ $h^3$\MpcV\ at $z\sim0.5$. The stellar mass function evolves very little in this redshift range suggesting that CMASS galaxies are passive and show almost or no on-going star formation \citep{Maraston13}. A non-evolving sample of massive galaxies provides an excellent ``cosmic laboratory'' to study galaxy formation and evolution as shown by \citet[][]{Bernadi16,Montero-Dorta16,Montero-Dorta17}, and their link to cosmology via e.g. the large-scale structure distribution and clustering of \boss galaxies studied by \citet[][]{Chuang16,Rodriguez-Torres16,Guo18}. \boss LRGs were repeatedly used to determine fundamental cosmological parameters: \citet[][]{Cuesta16,Gil-Marin17,Ross17} and to put cosmological models to the test: e.g.  \citet[][]{Anderson14,Beutler14_neutrino,Alam17_cosmoAnal,Sullivan17,Mueller18}. Furthermore, because the sample addresses the most luminous and red galaxies, they act as an important probe to close the gap in understanding the link between dark matter haloes and massive galaxies \citep[][]{Leauthaud12_HOD,Nuza13_HAM,Guo14_HOD,Saito16,Favole16b}. \par

At low redshift LRGs are known to populate the most massive haloes located in denser regions such as the centre of clusters and superclusters \citep{Lietzen12}. That makes them particularly interesting to study, because they give clues to the assembly of the most massive structures, the formation of haloes, and their connection to their associated galaxies. Thereby the ratio of their stellar to halo masses as a function of halo mass (\SHMF) allows for exploring the galaxy-halo connection and the formation and evolution of those galaxies in dark matter haloes of a certain mass range. Or equally, what halo mass is related to a galaxy that produced a certain stellar mass over a certain time. From a more cosmological point of view the relation shows how galaxies trace dark matter and how its density field is distributed\footnote{From the density field the corresponding power spectrum can be constructed and from that cosmological parameter determined. One can see that this simple relation between stellar and halo mass is indeed a powerful constraint.}. Interestingly, the haloes at intermediate masses produce stars most efficiently, relative to their mass \citep{White&Frenk91,Benson03,Bower06}. It is still barely understood why haloes with lower or higher masses are by orders of magnitudes less efficient \citep{Behroozi13c}. To shed light on this topic one would need to study the full history of mass assembly and star formation within a large redshift range, which is a costly task for ``full-physics'' hydro-cosmological simulations. The number of particles in question to cover a similar physical volume and amount of galaxies as an observational survey is therefore inaccessible. Different approaches to modelling the population of dark matter haloes with galaxies as well as their formation and evolution inside the haloes, have been developed. One of them being \textit{semi-analytical models} (hereafter SAMs). SAMs are usually build upon N-body dark matter simulations (e.g \textsc{Millennium}: \citet{Springel05_MSI}, \MD: \citet{Klypin16_MD} using merger trees (information of the hierarchical formation of dark matter haloes) and implementing baryonic physics as a post-processing step. For details on semi-analytical modelling we refer to excellent reviews on the field: \citet[][]{Baugh06,Benson10,Baugh12,Somerville&Dave15,Cora16}.\par

SAMs have been used recently in various frameworks to study for example correlation functions and galaxy clustering \citep[][]{Campbell15,Farrow15,vanDaalen16}, the galaxy-halo connection \citep[][]{Contreras13,Contreras15}; or active galactic nuclei, galaxy mergers, and the cosmic web \citep[][]{Almeida08,Lui16_LRGs,Ren18,Shirakata18}. They have been utilised to trace the star formation history \citep[][]{Mutch13,Lagos14,Orsi14,Gruppioni15}; to understand the galaxy mass-luminosity relations \citep[][]{Zoldan18}; or the processes regulating star formation \citep{Henriques17,Henriques18,Cora18b} or generating galaxy colours and metallicities \citep[][]{Yates12,Gonzalez-Perez14,Rodrigues17_GALFORM,Xie17,Collacchioni18}.\par

Within this paper we connect two major frameworks using a SAM: \textit{galaxy clustering} and \textit{galaxy formation}, in order to learn about the nature and properties of those most massive galaxies. \citet{Contreras13} performed a similar work and claimed that galaxy properties, apart from the stellar mass, e.g., star formation rate or cold gas mass, have more complicated correlation and non-negligible impacts on the clustering. Thereby the type of galaxy (central or satellite) plays a crucial role. \citet{Knebe17_MD} did a similar study with the \MD-SAMs for the \sdss main sample ($z \sim 0.1$). Within our work we expand upon these studies focusing at the redshift $z\sim0.5$ and \cmass galaxies. For that we use the same publicly available galaxy catalogues called the ``\MDG''. From them we take the SAM-code \galacticus as our modelled galaxy catalogue because it provides proper luminosities in the \textsc{SDSS} $ugriz$-band magnitudes suitable to compare with data from \boss (Data Release 12), which we adopt as our observational sample. \par

This paper is organised as the following: In \hyperref[sec:samples]{\Sec{sec:samples}} we describe the observational and modelled galaxy samples. In \hyperref[sec:modelling]{\Sec{sec:modelling}}  we show how to replicate the \cmass photometric selection for our model, \galacticus. We further provide confidence plots and a detailed study of various galaxy properties in \hyperref[sec:comp]{\Sec{sec:comp}}. Our results and discussion can be found in \hyperref[sec:results]{\Sec{sec:results}} and \hyperref[sec:discussion]{\Sec{sec:discussion}}, respectively, and our summary in \hyperref[sec:summary]{\Sec{sec:summary}}. The adopted cosmology in the \MDG as well as in this paper consists of a flat $\Lambda$CDM model with the following cosmological parameters: $\Omega_m=0.307, \Omega_b=0.048, \Omega_\Lambda=0.693, \sigma_8=0.823, n_s=0.96$, and a dimensionless Hubble parameter $h=0.678$ \citep{Planck15}. Hereafter, $h$ will is absorbed in the numerical value of its property throughout the text and in all tables and figures.

\section{Data Sets and Selection} \label{sec:samples}

We use \boss-\cmass galaxies as our observational and the semi-analytical \MDgal galaxy catalogue product as our modelled data sample. In this section we show the selection algorithms used to generate those samples. We further document all necessary assumptions and corrections applied to the samples in order to create comparable observational and modelled data sets. Those corrections include e.g. adjusting galaxy properties to our chosen cosmology (observations) or generating colours from luminosities (model).

\subsection{Observational Data: The \boss-\cmass Sample} \label{sec:data}
The \cmass-sample was designed to target the most luminous red galaxies in order to produce a uniformly (in mass) distributed samples of galaxies at redshift $0.43<z<0.7$ by applying a set of colour-magnitude cuts \hyperref[eq:cut_dmesa]{\Eq{eq:cut_dmesa}-\EqO{eq:dmesa_def}} shown below. The \cmass selection is similar to the algorithms used to target \texttt{SDSS-I/II Cut-II} \citep{Eisenstein01_LRGs} and \texttt{2SLAQ LRGs} \citep{Cannon06_LGRs}, using \gr and \ri colours to isolate high redshift galaxies, but the algorithm guarantees for an extension towards the bluer colours and the so called ``blue-cloud'' (BC) galaxies can enter the \cmass-sample. In our study we use \boss data from Data Release 12 \citep[hereafter \bossDR;][]{Alam15_SDSS_DR12}. The following colour-magnitude cuts are used to select the \cmass galaxies: 

\vspace{-0.4cm}\begin{equation}
 d_{\perp} > 0.55, \label{eq:cut_dmesa}
\end{equation} \vspace{-0.8cm}	
\begin{equation}
 i < 19.86 + 1.6~(d_{\perp} - 0.8), \label{eq:cut_dmesa_lt_i}
\end{equation} \vspace{-0.8cm}	
\begin{equation}
 17.5 < i < 19.9, \label{eq:cut_i}
\end{equation} \vspace{-0.8cm}	
\begin{equation}	
 r-i  < 2, \label{eq:cut_r-i}
\end{equation}	\vspace{-0.8cm}
\begin{equation} 		
 i_{\rm fib2} < 21.5, \label{eq:cut_ifib2}
\end{equation} \vspace{-0.8cm}
\begin{equation}	
 i_{\rm psf} - i_{\rm mod} > 0.2 + 0.2~(20.0 - i_{\rm mod}), \label{eq:cut_ipsf-imod}
\end{equation} \vspace{-0.8cm}
\begin{equation}	
 z_{\rm psf} - z_{\rm mod} > 9.125 - 0.46~z_{\rm mod}, \label{eq:cut_z}
\end{equation}
\noindent where \dperp is called the ``composite colour'' with: \\
\vspace{-0.4cm}\begin{equation}
	  d_{\perp} = (r - i) - (g - r) / 8.0, \label{eq:dmesa_def}
\end{equation}

\noindent $g,r,i$ are the \textit{cmodel} magnitudes in the AB-system, $i_{\rm mod}$ and $z_{\rm mod}$ refer to \textit{model} magnitudes, $i_{\rm fib2}$ is the \textit{fiber} magnitude, and $i_{\rm psf}$ and $z_{\rm psf}$ are the \textit{PSF} magnitudes. For more information about the set of colour-magnitudes cuts consult the \bossDR target selection webpage\footnote{\url{http://www.sdss.org/dr12/algorithms/boss_galaxy_ts/}}. \hyperref[eq:cut_dmesa]{\Eq{eq:cut_dmesa}} isolates high-redshift objects; \hyperref[eq:cut_dmesa_lt_i]{\Eq{eq:cut_dmesa_lt_i}} is a sliding magnitude cut that selects the brightest or more massive galaxies with redshift; \hyperref[eq:cut_i]{\Eq{eq:cut_i}} defines the faint and bright limits; and \hyperref[eq:cut_r-i]{\Eq{eq:cut_r-i}} protects from some outliers. \hyperref[eq:cut_ifib2]{\Eq{eq:cut_ifib2}} ensures a high redshift measurement success rate; and \hyperref[eq:cut_ipsf-imod]{\Eq{eq:cut_ipsf-imod}} and \hyperref[eq:cut_z]{\Eq{eq:cut_z}} perform a star-galaxy separation.\par

We use the latest ``Large-Scale Structure (LSS) catalogue''\footnote{\url{https://data.sdss.org/sas/dr12/boss/lss/}} \citep{Reid16_BOSS_DR12_LSS} from the \textit{SDSS Science Archive Server} which was cross-matched with the Portsmouth\footnote{\url{http://www.sdss.org/dr13/spectro/galaxy_portsmouth/}} passive galaxy sample to include stellar masses. The stellar masses were generated via a post-processing step using the stellar population models of \citet{Maraston05} and \citet{Maraston09} to perform a best-fit to observed \textit{ugriz}-magnitudes \citep{Fukugita96_ugriz}. \par

We use \texttt{Planck} cosmology and assume a \citet{Chabrier03} initial mass function (IMF). The Portsmouth galaxy product assumes a \texttt{WMAP7} flat $\Lambda$CDM cosmology with a dimensionless Hubble parameter of $h=0.7$ \citep[][same as in the entire \boss pipeline]{White11} and a \citet{Kroupa01} IMF. Therefore we correct their stellar masses from \texttt{WMAP7} to \texttt{Planck} cosmology\footnote{In order to translate between cosmologies we assume the simple relation of $\logT \frac{\Mstar^{\rm Planck}}{\Mstar^{\rm WMAP7}} \propto \logT \frac{D^{\rm WMAP7}_{\rm c}}{D^{\rm Planck}_{\rm c}}$, with \Mstar\ being the stellar mass and $D_{\rm c}$ the comoving distance within a certain cosmology}. We further convert the stellar masses to match the assumed IMF of \MDG models, \citet{Chabrier03}, with the following conversion: $\logT M_{\rm Chabrier}=\logT M_{\rm Kroupa}-0.03925$ \citep[see Table B1 in][]{Lacey16}. \par

For the data reduction we use the same approach as \citet[][]{Rodriguez-Torres16}, described in their Sec.2. In order to account for redshift failure and fiber collision we apply weights given by \citet{Anderson14}, using Eq.(9) in \citet[][]{Rodriguez-Torres16}. This results in a total number of 818,817 observed \cmass galaxies (entire redshift range). For this work we select a sub-sample of galaxies in the range $0.5<z<0.6$, which guarantees for maximal completeness in number density \citep{Guo18}, leaving us with a catalogue of 423,671 galaxies to study. We use this selection to compute the stellar mass function and clustering of the observed galaxies using the \texttt{Planck} parameters as a fiducial cosmology. We also extract the bias and number density from this sample to construct a Halo Abundance Matching (HAM) on the \bmd simulation that describes these observations. Furthermore, the BOSS survey covers around $\sim 9,600$ $\rm deg^2$ of the sky which corresponds to a volume of $\sim 4.147 \times 10^9~\rm Mpc^3$ within our redshift range and assumed cosmology.

\subsection{\MDG: \MDgal}\label{sec:galacticus}
\MDgal is based on the semi-analytical galaxy formation and evolution code \galacticus from \citet{Benson12} and consists of a large catalogue\footnote{The galaxy catalogue is publicly available on \url{www.cosmosim.org} and \url{www.skiesanduniverses.org}.} of galaxy properties including the \textsc{SDSS} $ugriz$-band luminosities. It was run on the 1000\hMpc\ dark matter simulation \textsc{MultiDark Planck 2} \citep[hereafter \MDPL:][]{Klypin16_MD} following the evolution of $3840^3$ dark matter particles with a mass per particle of $m_p=2.23 \times 10^9$ \Msun and minimum 20 particles/halo. Haloes and sub-haloes were identified with \rockstar\ \citep[][]{Behroozi13a} and merger trees constructed with \consistenttree\ \citep[][]{Behroozi13b}. The \galacticus SAM assumes a stellar population synthesis model from \citet{Conroy09} and a dust model of \citet{Ferrara99}.  The definition of the dark matter halo mass is giving by:

\vspace{-0.4cm}\begin{equation}
 M_{\rm ref}(<R_{\rm ref}) = \Delta_{\rm ref} \rho_{\rm c} \frac{4\pi}{3} R_{\rm ref}^3, \label{eq:mass_def}
\end{equation} \vspace{-0.4cm}

\noindent where $\Delta_{\rm ref}=\Delta_{\rm BN98}$ for \MBN\ with $\Delta_{\rm BN98}$ being the virial factor as given by the Eq. (6) of \citet{Bryan&Norman98}; $\rho_{\rm c}$ being the critical density of the Universe, and $R_{\rm ref}$ being the corresponding halo radius for which the interior mean density matches the desired value on the right-hand side of \hyperref[eq:mass_def]{\Eq{eq:mass_def}}. For information on the models' calibration and intrinsic constrains, we refer to the \MDG data release paper \citet[Sec.2.2 and Table 1]{Knebe17_MD}.\par

\galacticus\ returns luminosities, $L$, in the SDSS $ugriz$-bands at the zero-point of the AB-magnitude system in units of $4.4659\times 10^{13}\rm WHz^{-1}$. We apply $M_{AB}=-2.5\logT L$, to convert $L$ to absolute magnitudes $M_{AB}$ in each filter band. The filter band was by default blue-shifted to the redshift of the galaxy; meaning that in order to compute the apparent magnitude one must add not only the distance modulus, but also a factor of $-2.5\logT(1+z_0)$ to account for the compression of the photon frequencies at $z_0=0.56$. This results in

\vspace{-0.4cm}\begin{equation}
  m_{AB} = M_{AB} + DM(z) -2.5\logT(1+z_0), \label{eq:mAB}
\end{equation} \vspace{-0.4cm}

\noindent with $m_{AB}$ being the observed apparent magnitude in the AB-system and $DM(z)=5\logT(D^z_L/10\rm pc)$ the distance modulus with $D^z_L$ as luminosity distance at the redshift $z=0.56$ in parsec.

\section{Sample Selection and Colour-Magnitude Evaluation}\label{sec:modelling}
\begin{table*}
 \begin{center}
  \setlength{\tabcolsep}{3.5pt}
   \begin{tabular}{ll|r|rrr|cc|l}
      \hline
       data 	&sample name& \Ngal 	& f$^{\rm total}_{\rm c}$ & f$^{\rm total}_{\rm sats}$ & f$^{\rm sats}_{\rm o}$ & $n$ $\times 10^{-4}$ & \Vz $\times 10^9$ & remark \\
		& 		& total	& centrals		& total sats	& orphan sats& [\MpcV] & [\MpcCu] & \\
		& 		& 		& (\Ngal) 		& (\Ngal)		& (\Ngal)	&  &  & \\
      \hline
      \hline
      \bossDR& \cmDR	& 423,671	& $\sim$0.900	& $\sim$0.100	& - 		& 1.02 & 4.147 & $0.5<z<0.6$\\
      \hline
      \MDgal 	& \all		& 1,844,542& 0.794		& 0.206	 	& 0.205  	& 5.737 & 3.212 & entire set of galaxies\\
		& 		& 		& (1,465,070)	& (379,472) 	& (64,478)  	& & & $\Mstar>10^{10.7}$ \Msun\\
      \hline
      \MDgal 	& \cm	& 95,683	& 0.901		& 0.089	 	& 0.112  	& 0.30 & 3.212 & set of colour-magnitude\\
		& 		& 		& (87,167)		& (8,516) 		& (859)  	& & &  cuts: \hyperref[eq:cut_dmesa]{\Eq{eq:cut_dmesa}-\EqO{eq:cut_r-i}}\\
      \hline
      \MDgal 	& \den	& 314,083	& 0.848		& 0.151	 	& 0.171  	&  1.02  & 3.212 & red-blue cut using \citet[][Eq.7]{GuoH13}\\
		& 		& 		& (266,483)		& (47,600)		& (6,952)  	& & & down-sampled with \SMF at $z=0.56$ \\
      \hline
      \MDgal 	& \ma	& 129,109	& 0.899		& 0.101	 	& 0.118	& 0.40 & 3.212 & $\Mstar>10^{11.24}$ \Msun\\
		& 		& 		& (116,120)		& (12,989)		& (1,373) 	& & &\\
      \hline
      \hline
	(i)	& 	(ii)	& 	(iii)	& (iv)			& (v)			& (vi) 	& (vii) & (viii)& (ix)\\
      \hline
    \end{tabular}
    \caption{The table summarises the properties of the observed and modelled galaxy samples used in our study. Column (i) shows the name of the publicly available galaxy catalogue we extracted a sample form, (ii) gives the label of the corresponding sample throughout this paper, and (iii) its total number of galaxies $\Ngal$. The corresponding fraction of central, satellite or orphan galaxies can be found in (iv) f$^{\rm total}_{\rm c}$ for centrals, (v) f$^{\rm total}_{\rm sats}$ for all satellites (non-orphans+orphans), and (vi) f$^{\rm sats}_{\rm o}$ for orphan satellites (the fraction of orphan satellites is calculated with respect to the total number of satellites), respectively. The number density $n$ of each sample and the effective volume \Vz\ can be found in Column (vii) and (viii), respectively. Column (ix) provides comments on the selection. For the observational sample we select \bossDR galaxies in the redshift range of $0.5<z<0.6$ and label the sample \cmDR. For the modelled galaxies we show the entire galaxies sample above a confidence cut in stellar mass of $\Mstar>10^{10.7}$ \Msun: \all and the following CMASS-mock samples: \cm, \den, and \ma at redshift $z=0.56$ (which matches the median redshift of the full \cmass sample). To extract \cm the standard set of \cmass colour-magnitude cuts from \hyperref[eq:cut_dmesa]{\Eq{eq:cut_dmesa}-\EqO{eq:cut_r-i}} was applied. For \den\ we used a down sampling algorithm shown in \hyperref[eq:Guo13]{\Eq{eq:Guo13}} and \hyperref[eq:down]{\Eq{eq:down}}, where we selected randomly galaxies from the red population that matched the number density of \cmDR. For \ma a stellar mass cut at $Mstar>10^{11.24}$ \Msun\ was applied according to the findings of \citet{Maraston13}.}
    \vspace{-0.4cm}\label{tab:tarsel}
  \end{center}
\end{table*}

In this section we show how we extracted a CMASS-mock sample from the \MDgal catalogue. Since we deal with modelled galaxy properties we only use a limited set of colour-magnitude selection cuts, \hyperref[eq:cut_dmesa]{\Eqs{eq:cut_dmesa}-\EqO{eq:cut_r-i}}, because the simulation does not distinguish between \texttt{model} and \texttt{cmodel} magnitudes\footnote{``\texttt{model}'' and ``\texttt{cmodel}'' refer to different approaches of how magnitudes have been generated through the photometric pipeline of \texttt{SDSS}.}.\par

In order to test our CMASS-mock samples we compare on the one hand to observed \cmass galaxies from the \textit{Portsmouth} merged galaxy catalogue of the $12^{th}$ data release (referred to as \cmDR) in the redshift range of  $0.5<z<0.6$ (the most complete range in terms of stellar masses), which corresponds to a comoving number density of $n=1.02 \times 10^{-4}$ \MpcV at redshift $z\sim0.55$ in our adopted cosmology. And on the other hand we extract two more CMASS-mock samples aiming at reproducing the colour-magnitude selection by using other galaxy properties as stellar mass. We do that because luminosities or colours are not always available for modelled galaxy samples, especially if they are as large as \MDPL. Furthermore, we can assess the colours and luminosities of our SAM by comparing it with a sample selected by applying a high stellar mass cut. Both methods should produce similar catalogues, because we expect that the most massive galaxies and the brightest and reddest galaxies coincide with each other. \par

Therefore we create a second and a third CMASS-mock sample by matching the number density and the stellar mass distribution of the observed sample \cmDR, or by applying a high stellar mass cut corresponding to \cmass galaxies as reported by \citet{Maraston13}, respectively. We summarise our sample selection in the following list:

\begin{itemize}
  	\item[] \noindent \textbf{\all:} resulting full sample of $\sim 1.8 \times 10^6$ galaxies after applying a confidence cut in stellar masses\footnote{This stellar mass threshold correspond to a conservative confidence cut above the output of the model can be trusted -- see \MDG release paper for details}: $\Mstar>10^{9.5}$ \Msun; this is the entire sample of \galacticus\ at $z=0.56$
  	\vspace{0.2cm}

  	\item[] \noindent \textbf{\cm:} colour-selected sample; the observational \cmass colour-magnitude selection, \hyperref[eq:cut_dmesa]{\Eqs{eq:cut_dmesa}-\EqO{eq:cut_r-i}}, described in \hyperref[sec:data]{\Sec{sec:data}}, has been applied\footnote{We use dust-extincted luminosities in our study because we compare with observations. If we would use non-dust corrected luminosities instead, we would find very small differences of about $\Delta_{MAB_{gri}} \sim 0.1-0.2$ mags in $gri$-bands compared to dust-extincted magnitudes.}
	\vspace{0.2cm}

  	\item[] \noindent \textbf{\den:} number density-selected sample; the number density of \bossDR ($n_{\rm \cmass}=1.02 \times 10^{-4}$ \MpcV) was matched via randomly down-sampling the red population of \allS stellar mass function (\SMF) at $z=0.56$. The red population was selected with a cut in colour as introduced by \citet[][Eq.7]{GuoH13}:
  	\begin{equation}
	r-i>0.679-0.082~(M_i-20). \label{eq:Guo13}
  	\end{equation}
 
  	\noindent We use \hyperref[eq:Guo13]{\Eq{eq:Guo13}} instead of a simple cut in red-blue separation as $\gi>2.35$ because otherwise we would exclude a significant amount of galaxies at $\Mstar\sim10^{11.2}$ \Msun\ and fail to calculate the true stellar mass function. After applying the colour selection, we calculate the fraction between the densities of the stellar mass functions $\Phi$ $\rm dex^{-1}\MpcV$ of \cmDR and \galacticus\ and use it to compare to a random distribution, $S_{\rm rand}$, between $[0,1)$:
  	\begin{equation}
  		S_{\rm rand} < \frac{\Phi_{\cmDR}}{\Phi_{\galacticus}}, \label{eq:down}
  	\end{equation}  
  	\noindent A galaxy enters the sample if the condition in \hyperref[eq:Guo13]{\Eqs{eq:Guo13}} is fulfilled, otherwise it is discarded.
   	\vspace{0.2cm}
   
  	\item[] \noindent \textbf{\ma:} stellar mass-selected sample; we apply a stellar mass $\Mstar>10^{11.24}$ \Msun\ on \all (see \citealt{Maraston13}) 
\end{itemize}

\noindent In \hyperref[tab:tarsel]{\Tab{tab:tarsel}} we summarise the properties of our observational and modelled CMASS samples. We show the total number of galaxies \Ngal, total numbers and fractions of ``centrals'', ``satellites'', and ``orphan (satellites)''\footnote{``Orphan'' or ``orphan satellite'' is a technical term in semi-analytical modelling, referring to satellites which lost their dark matter haloes due to the interaction with their central galaxies or other reasons such as resolution limits of the halo finder.}, number densities $n$, and effective volumes \Vz. Although the \Ngal\ and $n$ are different in each CMASS-mock sample, the fraction of centrals (f$^{\rm total}_{\rm c} \sim 0.9$) and satellites (f$^{\rm total}_{\rm sats} \sim 0.1$) are almost identical and agree perfectly with the observation \citep{Guo14_HOD,Rodriguez-Torres16}. However, we note that the number density of the \cmS $n_{\rm \galacticus}=0.30 \times 10^{-4}$ \MpcV\ roughly corresponds to only 1/3 of the \bossDR with $\sim 1.02 \times 10^{-4}$ \MpcV. The discrepancy in the numbers and its consequences will be discussed later. In the following section we perform sanity checks on our \cm CMASS-mock by directly comparing with \bossDR data. Note that to avoid crowding we only show \cm and the observational sample in the figures.
 
\subsection{\cm: The Composite Colour \dperp}\label{sec:dperp_i}
\begin{figure}
  \includegraphics[width=8cm]{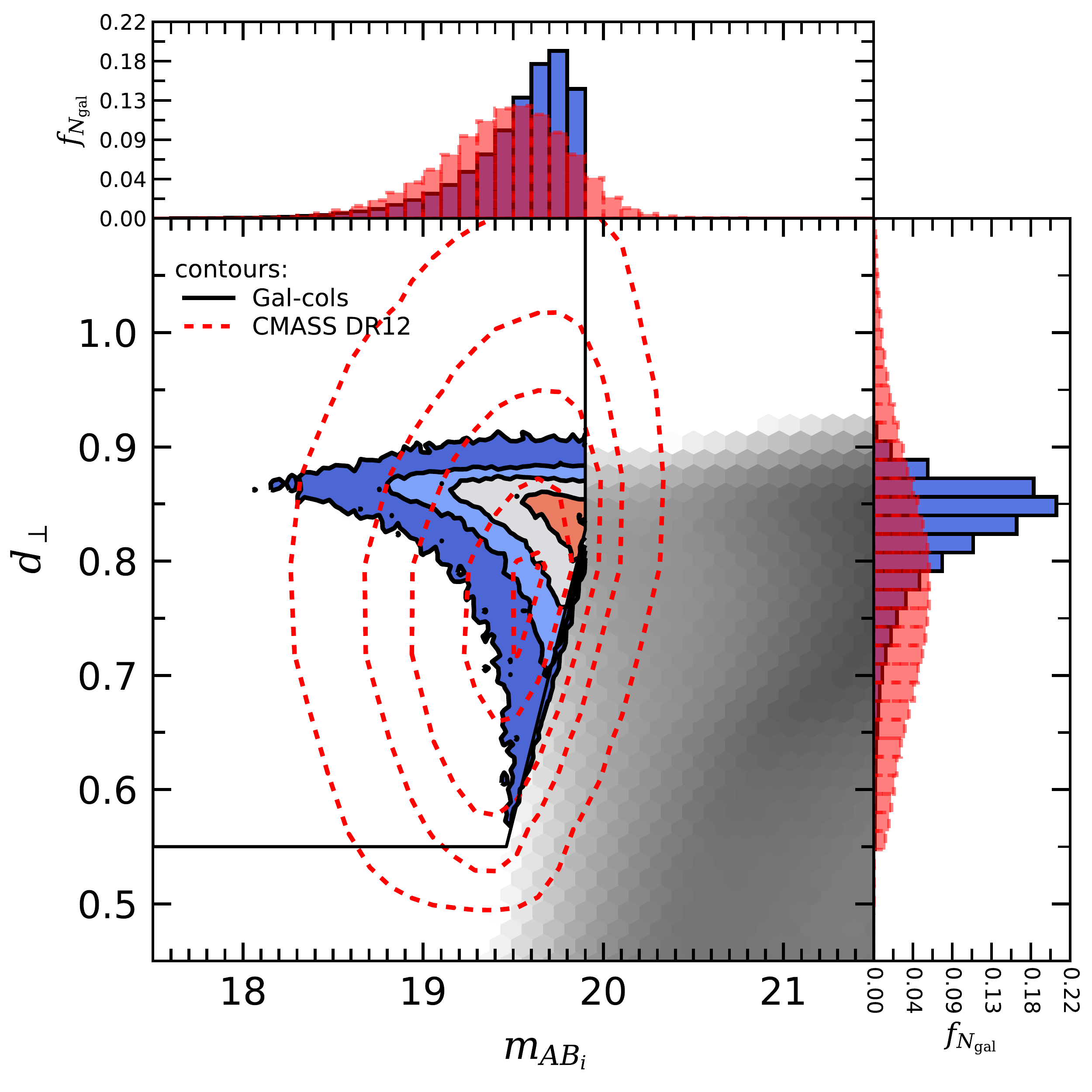}\vspace{-0.3cm} 
  \caption{Colour-Magnitude diagram for the modelled sample \cm (filled coloured contours) at $z=0.56$ and \bossDR galaxies in the range of $0.5<z<0.6$ (red dashed contours) for observed frame \dperp colour compared to observed apparent \iband magnitudes $m_{\rm AB_i}$. The solid black polygon-shaped area represents the \cmass colour-magnitudes cuts, the grey hexagons represent the total population of galaxies, \all. Modelled and observed galaxies are in very good agreement with each other.}\label{fig:dperp_i}
  \vspace{-0.4cm}
\end{figure}

The composite colour \dperp is a colour combination defined in \hyperref[eq:dmesa_def]{\Eq{eq:dmesa_def}} and the \textit{key colour selection parameter} for \cmass galaxies involving three bands: \textit{g,i}, and \textit{r}. \hyperref[fig:dperp_i]{\Fig{fig:dperp_i}} presents the colour-magnitude diagram (CMD) where \dperp\ is shown compared to the observed \textit{i}-band magnitudes, $m_{\rm AB_i}$. This is the first and most important sanity check we use to assess our colour selection. The CMASS colour-magnitude selection described in \hyperref[eq:cut_dmesa]{\Eq{eq:cut_dmesa}} and \hyperref[eq:cut_i]{\Eq{eq:cut_i}} are shown as a polygon-shaped area with a thin solid black line, where all galaxies within this area enter the selection. The \galacticus \cmass sample, \cm, is shown in black filled coloured contours and \bossDR in red dashed empty contours. We show the parameter space of the entire set of galaxies, \all, as grey logarithmic binned hexagons in the background to point out that the CMASS sample is only a tiny fraction of the total set of galaxies that \galacticus provides. For the contour-figures we use throughout this work the following confidence levels in per cent: [2.1, 13.6, 31.74, 68.26, 95, 99.7]. 

The histogram panels on the top and on the right hand side give information about the distribution of galaxies along the binned axes using 40 bins normalised by the total number of galaxies of each sample. The histograms show the same colour and line style keys as the contours: black solid lines and blue filled bars for \galacticus \cmS and red dashed lines and empty bars for \bossDR. The histogram of \all is not shown for reasons of over-crowding.

While the majority of the modelled galaxies lies outside the CMASS selection, we nevertheless report that a substantial number enter it. Their numbers can be found in \hyperref[tab:tarsel]{\Tab{tab:tarsel}} under the label \cm. We like to remark that \citet[][Fig.17]{Maraston13} report similar results for their adopted SAM. One can see in the histogram panels that \galacticus' number of galaxies in each bin is in general higher and less spread across the axes compared to the observations. In the next section we will discuss this issue in form of a colour-colour diagram in more detail.

\subsection{\cm: Colour-colour and Colour-mass Diagrams}\label{sec:r-i_g-r}
\begin{figure}
  \begin{center}
    \includegraphics[width=8cm]{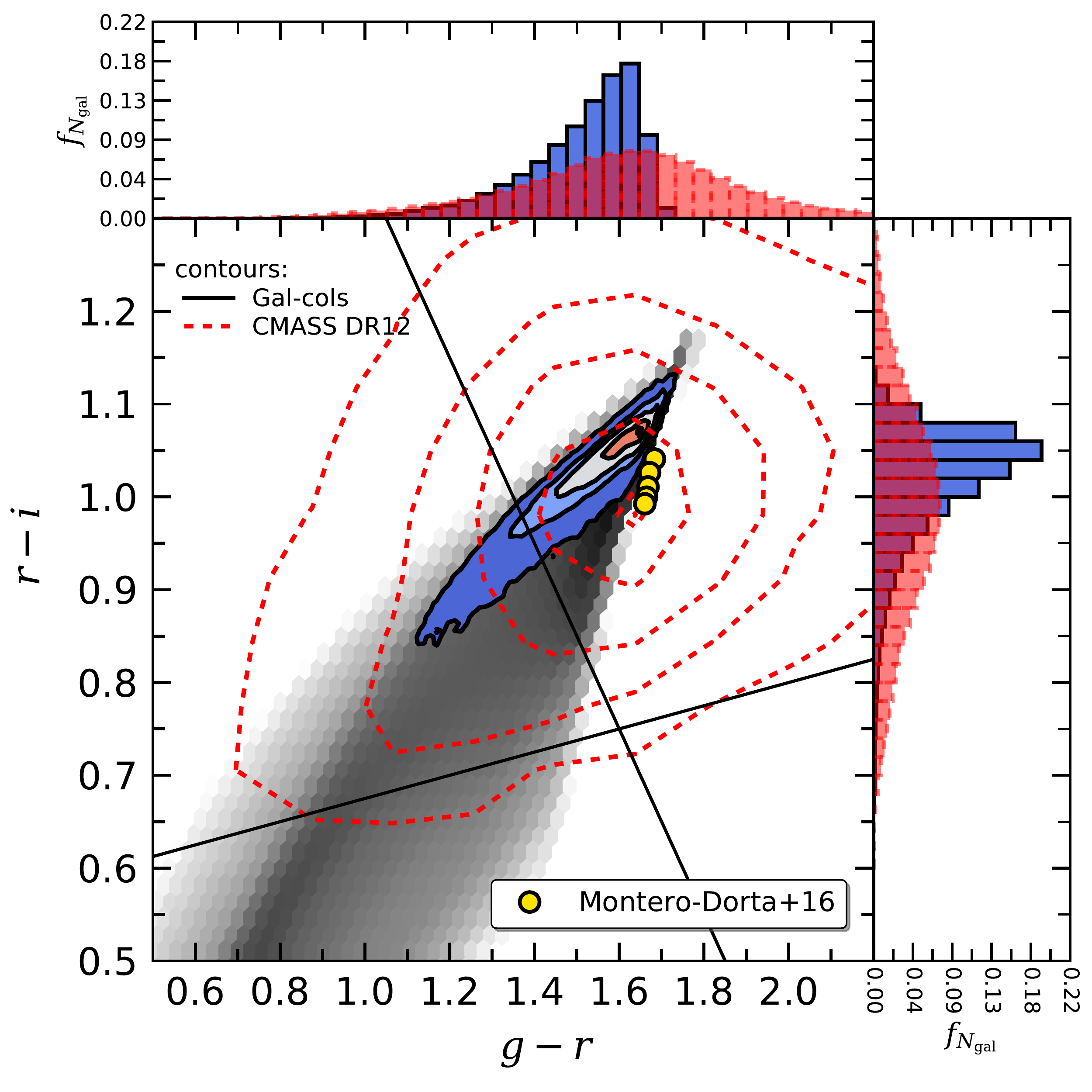}
    \includegraphics[width=8cm]{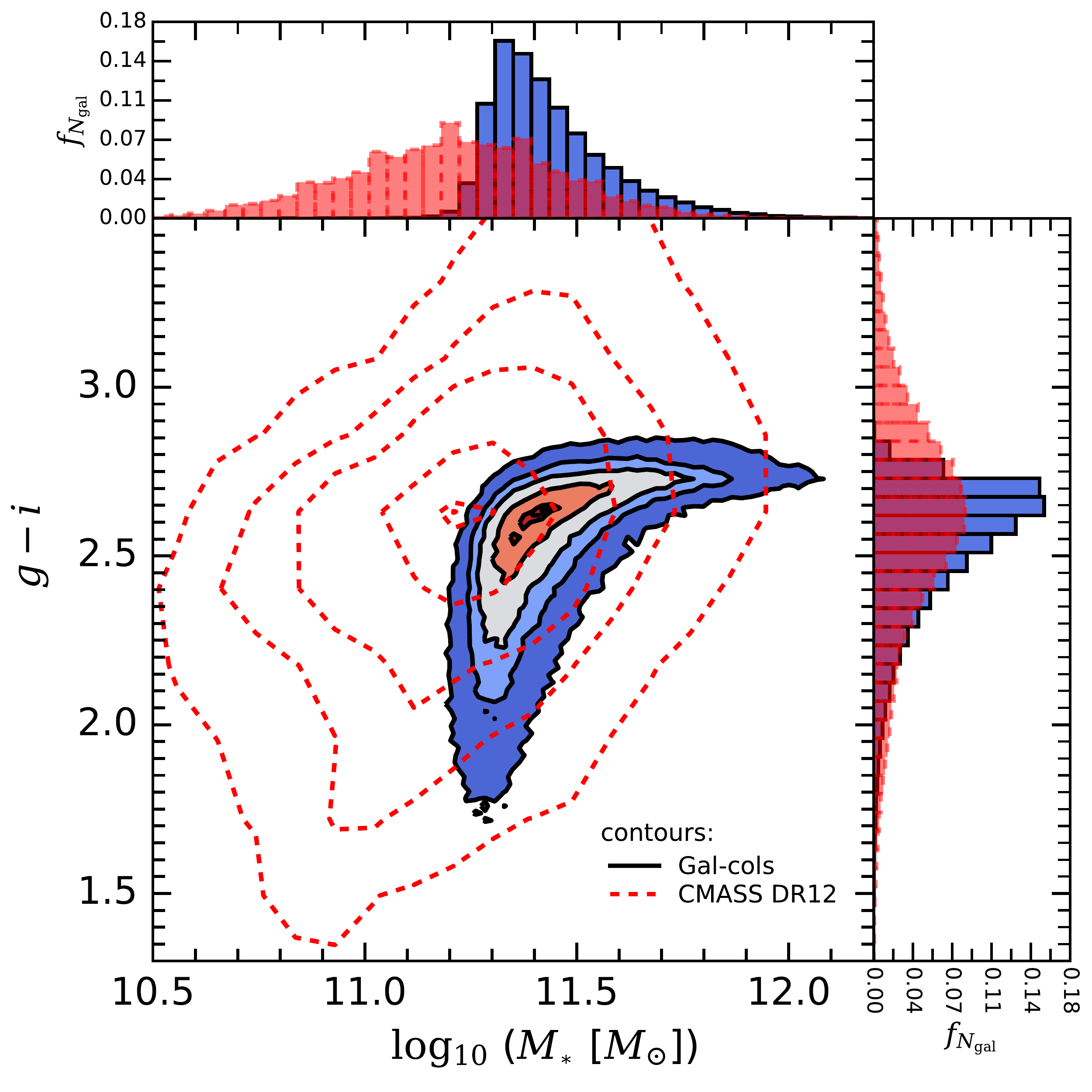}\vspace{-0.3cm}%
    \caption{\textit{Top:} Colour-colour diagram for observed colours \ri vs. \gr for \cm (filled coloured contours) and \cmDR (red dashed contours). The horizontal thin solid black line represents the \dperp-cut and the vertical thin solid black line the red-blue separation of $\gi=2.35$. The filled yellow circles show modelled RS of different \textit{i}-band magnitude slices from \citet{Montero-Dorta16}. \textit{Bottom:} Observed frame colour separation \gi vs. \Mstar.}\label{fig:r-i_g-r}
  \end{center}
  \vspace{-0.4cm}
\end{figure}

We show in the \textit{upper} panel of \hyperref[fig:r-i_g-r]{\Fig{fig:r-i_g-r}}, the \ri vs. \gr colour-colour diagram. The observed CMASS data (referred to as \cmDR) extends over a much larger region in the \ri and \gr than \galacticus \cm. This is most likely due to the fact that uncertainties (i.e., photometric errors) are not implemented in the model, so no artificial blurring was produced compared to the observations. We also note that the centroid of the \cm distribution is located at slightly redder colours ($\ri \sim 1.05$ and $\gr \sim 1.7$) than those of the observations and the location of the intrinsic ``red sequence'' (RS)  from \citet[][]{Montero-Dorta16}. The intrinsic RS is the a narrow sequence of massive red galaxies modelled as an extended Gaussian and is constituted as the counterpart to the ``blue cloud'' which is a more heterogenous population consisting of galaxies with bluer colours \citet[][]{Montero-Dorta16}. We further include the composite colour \dperp-cut as a horizontal; and a common separation of red and blue galaxies, $\gi=2.35$ \citep{Masters11}, as a vertical thin solid black line.\par

We show in the \textit{lower} panel of \hyperref[fig:r-i_g-r]{\Fig{fig:r-i_g-r}} the \gi colour dependence on stellar mass. The \cm' galaxies are slightly more massive (0.2 dex) than their observational counterparts from the Portsmouth merged catalogue, but the samples are in very good agreement.

\section{Sample Comparison} \label{sec:comp}
Since luminosities are due to many uncertainties involved in the SPS fitting (see e.g. \citealt{Conroy09}) much more complicated to model than masses, SAMs often reproduce only SMFs to a certain degree. Observations need to go the other way: fluxes have been measured and stellar SED fitting performed to assume stellar masses \citep{Maraston06}. Usually a huge computational effort was brought forward to create luminosities for SAMs applied to volumes as large as \MD. Therefore we want to investigate \textit{the variation in our samples of selecting CMASS galaxies by colour (as done in observations) vs. by other galaxy properties as stellar mass} (as mentioned in the previous section), using the fiducial plots from \hyperref[sec:modelling]{\Sec{sec:modelling}} once again.\vspace{-0.3cm}

\begin{figure}
    \begin{center}
      \includegraphics[width=6.8cm]{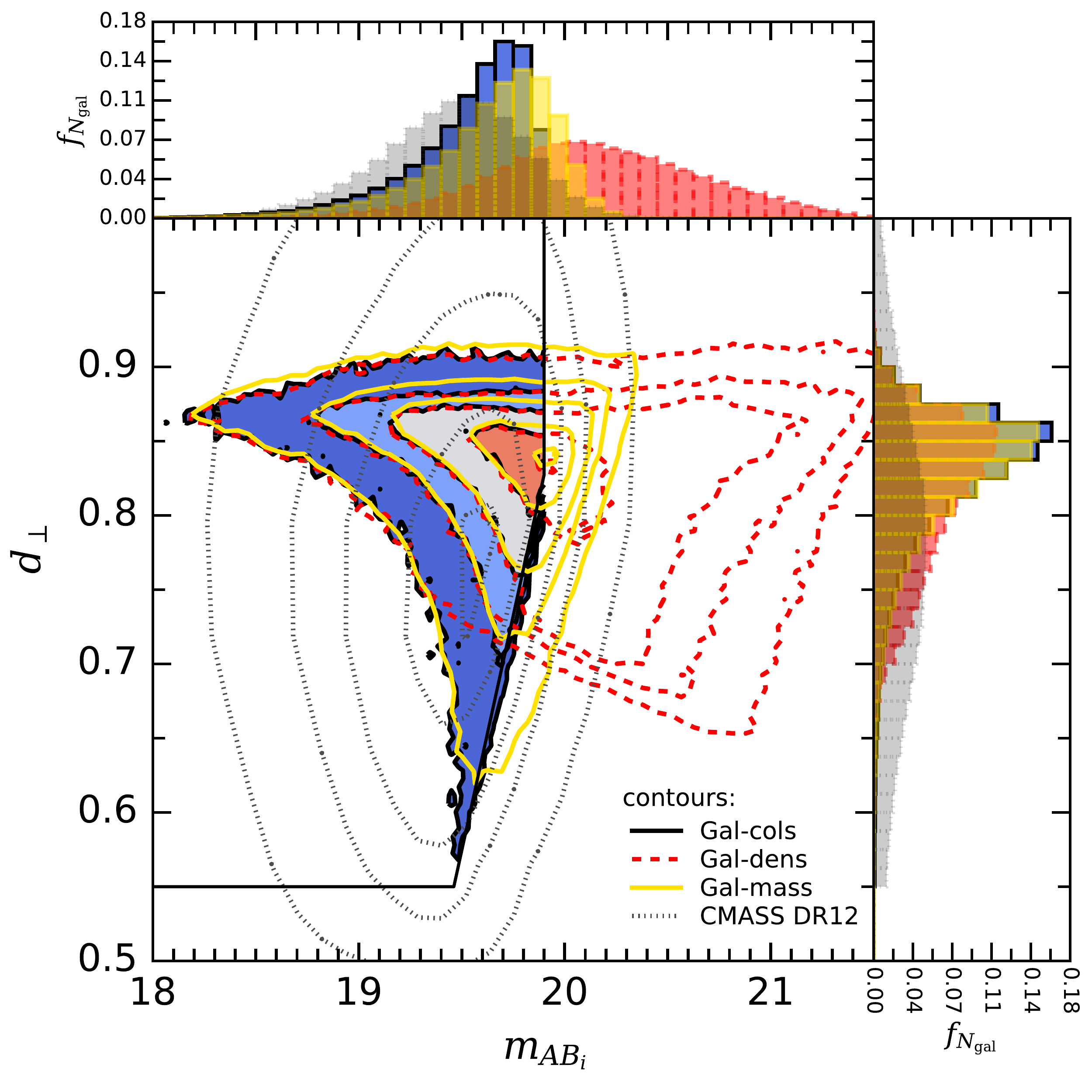}
      \includegraphics[width=6.8cm]{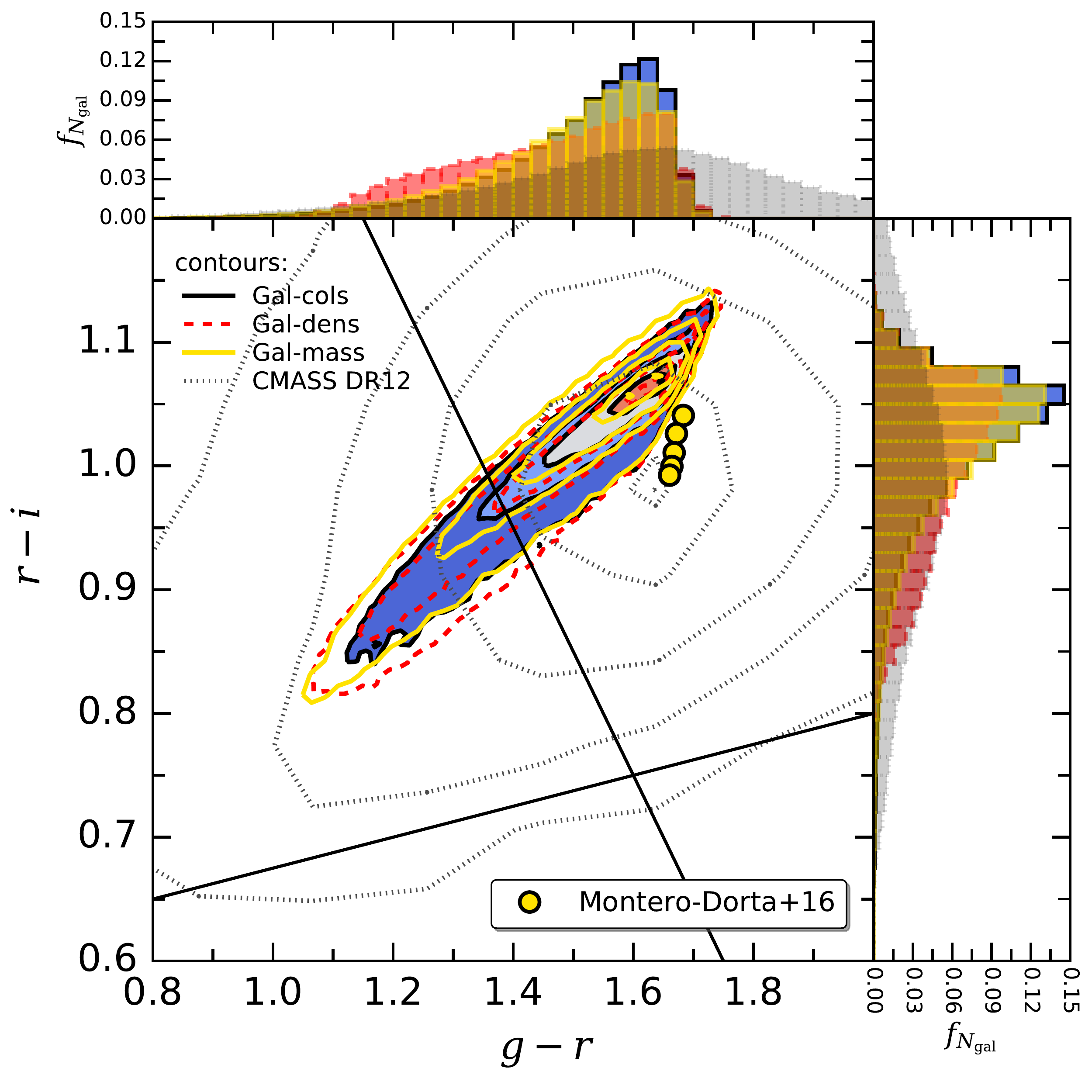}
      \includegraphics[width=6.8cm]{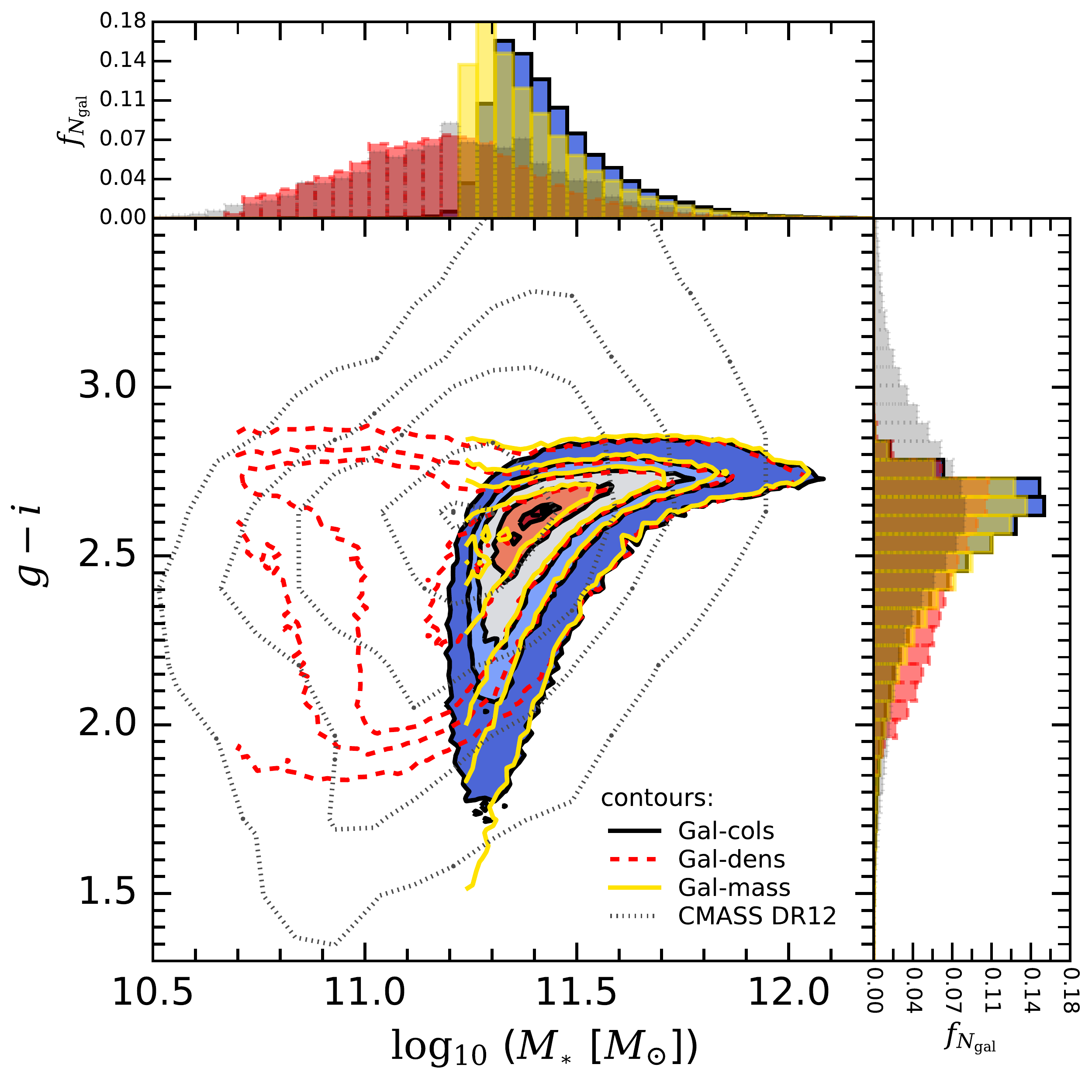}\vspace{-0.2cm}%
      \caption{Fiducial plots discussed in \hyperref[sec:modelling]{\Sec{sec:modelling}} including all three CMASS-mocks of \galacticus at redshift $z=0.56$: \cm (filled coloured contours), \den\ (red dashed contours), and \ma (yellow solid contours) compared to \cmDR  (dotted-dashed grey contours) within the range $0.5<z<0.6$. \textit{Top}: CMD. The solid black polygon-shaped area represents the \cmass colour-magnitude selection. \textit{Middle}: observed colour \ri vs \gr. The horizontal black line represents the \dperp-cut and the vertical black line in the same panel the red-blue separation of $\gi=2.35$. The filled yellow circles represent the modelled RS of different \iband magnitude slices from \citet{Montero-Dorta16}. \textit{Bottom}: \gi vs. \Mstar.}\label{fig:dis_fidplots}
    \end{center}
    \vspace{-0.4cm}
\end{figure}

\paragraph*{Colour-Magnitude diagram (CMD):}
\hyperref[fig:dis_fidplots]{\Fig{fig:dis_fidplots}} presents in the \textit{upper} panel the CMD (as in \hyperref[fig:dperp_i]{\Fig{fig:dperp_i}}) for the three modelled samples comparing observed frame \dperp\ colours to observed \iband magnitudes, $m_{\rm AB_i}$. A large part of the galaxies of the \denS\ and \maS lie outside the polygon reflecting the colour selection.The peak in magnitudes of \den\ is shifted 0.3 mags to fainter luminosities compared to \cm and extending into the low-luminosity regime. \ma agrees pretty well with \cm, where its peak is located exactly on the CMASS edge with $m_{\rm AB_i}=19.9$.\vspace{-0.3cm}

\paragraph*{Colour-colour diagram:}
In the \textit{middle} panel of \hyperref[fig:dis_fidplots]{\Fig{fig:dis_fidplots}} we show the colour-colour diagram for observed colours \ri vs. \gr (as in \hyperref[fig:r-i_g-r]{\Fig{fig:r-i_g-r}} \textit{lower} panel). The horizontal black line represents the \dperp-cut and the vertical black line the red-blue separation of $\gi=2.35$. The filled yellow circles show modelled RS of different \iband magnitude slices from \citet{Montero-Dorta16}. The three samples are in very good agreement with each others, but we can see that the galaxies of \den\ and \ma extend slightly toward ``bluer'' colours.\vspace{-0.3cm}

\paragraph*{Colour-mass diagram:}
In the \textit{lower} panel of \hyperref[fig:dis_fidplots]{\Fig{fig:dis_fidplots}} we show observed frame colour \gi vs. \Mstar\ (as in \hyperref[fig:r-i_g-r]{\Fig{fig:r-i_g-r}}, \textit{lower} panel). This figure shows that the mass distribution of the three samples is quite different. \den, which has the same number density as \boss, does not coincide with the sample selected by colour, \cm. However, the galaxies of the \den\ sample can be bound within the contours of \bossDR. Alternatively, a high-mass cut in stellar mass can be used to mimic the \cm sufficiently. The next paragraph is dedicated to studying the distribution of stellar masses in our samples in more detail.\vspace{-0.3cm}

\begin{figure}
  \includegraphics[width=8.4cm]{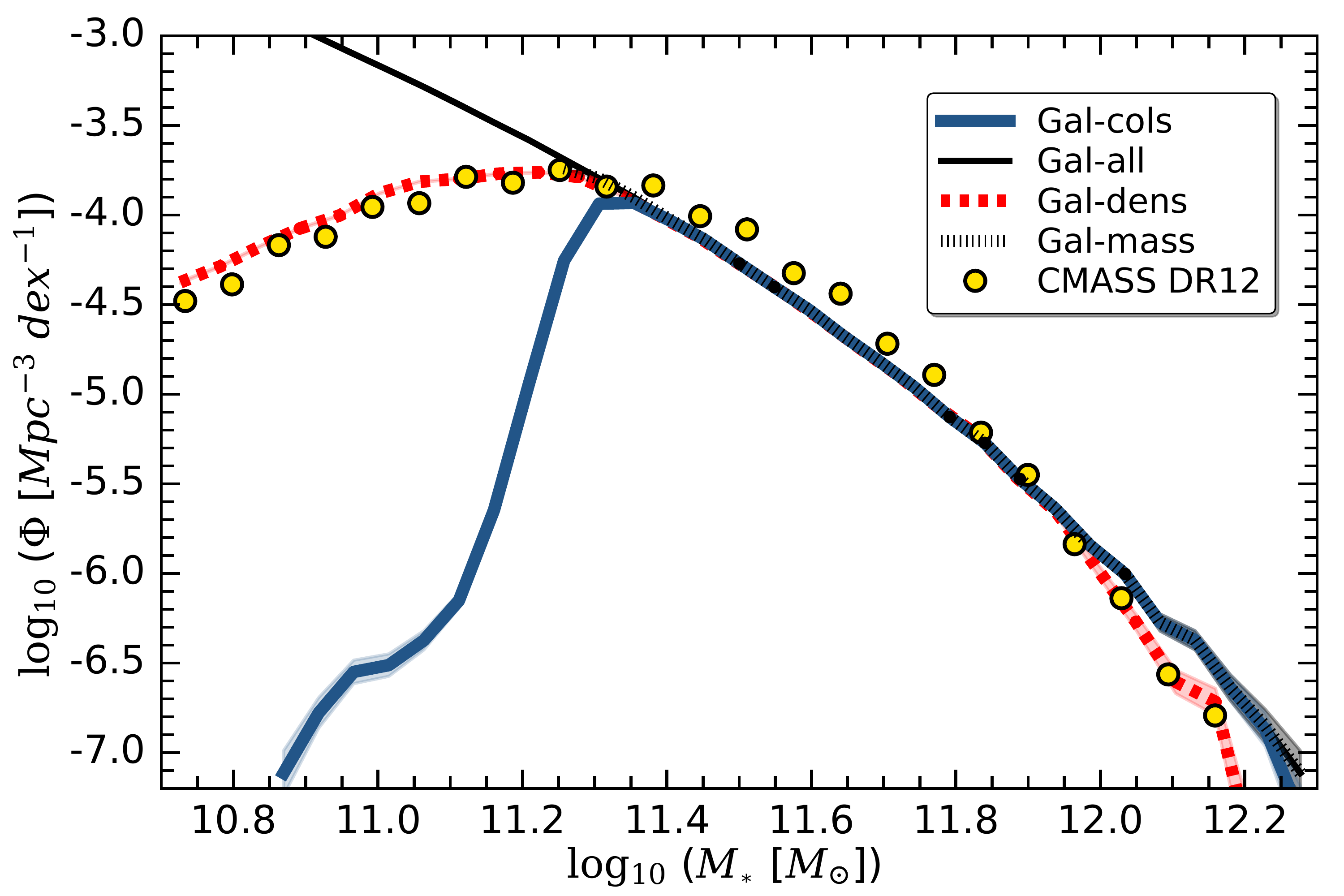}\vspace{-0.3cm}%
  \caption{\galacticus' stellar mass functions for the entire sample of galaxies (thin black line) and CMASS-mock samples: \cm- (blue solid line), \den\ (red dashed line), and \ma (grey dotted-dashed line) at redshift $z=0.56$ compared to \cmDR Portsmouth merged catalogue (filled yellow circles) in the range of $0.5<z<0.6$. Their errors bars are located within the size of the markers. In order to improve the readability of the figure, we removed the vertical line dropping to zero at $\Mstar>10^{11.24}$ \Msun\ due to the stellar mass cut applied on \ma.}\label{fig:smf}
  \vspace{-0.4cm}
\end{figure}

\paragraph*{Stellar mass function:} In \hyperref[fig:smf]{\Fig{fig:smf}} we present the stellar mass functions (\SMFs) at redshift $z=0.56$ for the total number of model galaxies from \galacticus \allS, as well as the CMASS-mocks: \cm, \ma, and \den\ compared to \cmDR (filled yellow circles). We state errors in the y-axis of the density functions as $\sigma_{\rm i} = \frac{y_{\rm i}}{\sqrt N_{\rm i}}$, where $\rm i = 0...n_{\rm bins}$, $y_i$ stands for the data on the y-axis, $N_{\rm i}$ for the number of galaxies in each bin, and $n_{\rm bins}$ for the number of bins.\par

As expected, the different CMASS-mock samples of \galacticus agree very well with each others. They show only slight variation at the high-mass end compared to \cm, due to the colour selection which excludes a few bright . Those could enter in \den\ and \ma because no colour selection was performed. At intermediate masses all three samples agree perfectly with each other, but their abundances lie slight beyond the observations. At lower masses we report that the \cmS shows the same typical shape of incompleteness in the stellar mass function as e.g. \citet[][Fig.3]{Rodriguez-Torres16} for the \textsc{BigMultiDark} \boss light-cone (\blc) or \citet{Maraston13}. \par

In summary we have shown that using a simple cut in stellar masses provides a good approximation for the observed \cmass sample. A number density sample (created with a down-sampling algorithm) draws the \SMF\ of \cmass perfectly, but permits bluer and low-mass objects to enter the sample. Those objects have fainter \iband magnitudes than \cmass as seen in \hyperref[fig:dis_fidplots]{\Fig{fig:dis_fidplots}} \textit{upper} panel. However, their colours and stellar masses are still in agreement with \cmass as shown in the \textit{middle} and \textit{lower} panel of \hyperref[fig:dis_fidplots]{\Fig{fig:dis_fidplots}}. In the following sections we will come back to the question if a CMASS-mock can be selected by other properties than colours and magnitudes and assess if a colour selections provides a more valid sample that a simple cut in stellar mass particularly for our SAM. Addressing a fully red population is crucial if one once to study CMASS galaxies, therefore we study the ``red sequence'' (RS) population and its \iband luminosity in the next paragraph.\vspace{-0.3cm}

\paragraph*{Luminosity function:}
\begin{figure}
  \includegraphics[width=8.4cm]{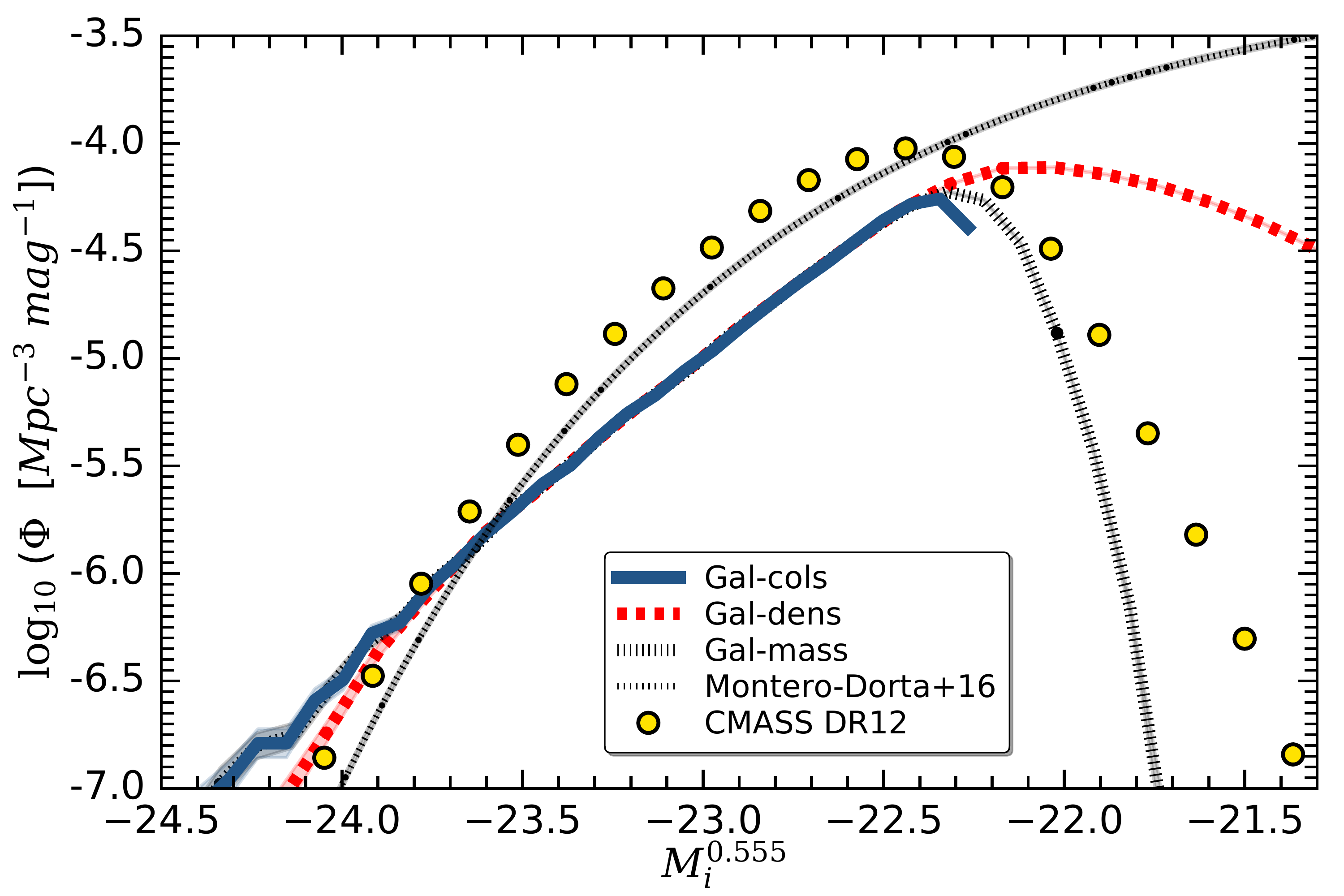}\vspace{-0.3cm}%
  \caption{Luminosity functions for \galacticus' CMASS-mock samples: \cm (blue solid line), \den\ (red dashed line), and \ma (grey dotted-dashed line) compared to the ``red sequence'' best-fit Schechter-function from \citet[][Table 3 \& Fig.14]{Montero-Dorta16} (thin grey dashed line) at redshift $z=0.555$. The errors of \bossDR are shown within the size of the markers.}\label{fig:RS}
  \vspace{-0.4cm}
\end{figure}

In \hyperref[sec:r-i_g-r]{\Sec{sec:r-i_g-r}} we briefly mentioned the ``red sequence'' (RS) population of CMASS galaxies. Now we want to discuss this topic in more detail and investigate if \galacticus' CMASS-mock galaxies also exhibit such a population. The RS can be found in observations as a sort of irregular blob in the \ri vs. \gr parameter space, elongated across the \gr-axis due to the \gband magnitudes higher error sensitivity. \citet{Montero-Dorta16} developed an analytic method to model the RS luminosity function (LF) and constrained Schechter-fit parameters. We mimic \galacticus' RS-samples by selecting red galaxies by applying \hyperref[eq:Guo13]{\Eq{eq:Guo13}} to \cm, \den, and \ma, respectively. In \hyperref[fig:RS]{\Fig{fig:RS}} we compare \galacticus' CMASS-mock samples to the best-fit of \citet{Montero-Dorta16} at $z=0.555$. The CMASS-mocks were further blue-shifted to the same redshift using an approximated K-correction of $-2.5\logT(1+z)$ \citep{Blanton&Roweis07} to fit the redshift of the Schechter-function\footnote{The fit uses \boss data which was deconvolved from photometric errors and selection effects  and show the raw, uncorrected, observed luminosity function. Photometric errors blur the colour-colour distribution (see the \textit{middle} panel in \hyperref[fig:dis_fidplots]{\Fig{fig:dis_fidplots}}), therefore objects scatter in and out the selection boundaries leading to the observed disagreement between the results of \cmDR (filled yellow circles) and the intrinsic red-sequence from \citet[][thin grey dashed line]{Montero-Dorta16}.}. 

We report that the reddest galaxies of \galacticus exceed the LF of the observations and the Schechter-fit of about 0.40 and 0.25 mags, respectively, at the bright end. At the faint end all three CMASS-mock samples poorly reproduce the Schechter-fit and their LF can roughly be estimated by a power law. We note that due to the cut in \iband magnitude (see \hyperref[eq:cut_i]{\Eq{eq:cut_i}}) \cm's LF is abruptly cut off at $M_{\rm AB}\sim-22.2$.

\section{Results} \label{sec:results}
In this section we present our results for the CMASS-mock samples \cm, \ma, and \den\ of \galacticus. We show stellar to halo mass ratios as a function of halo mass (\SHMFs), halo occupation distributions (HODs), and projected 2-point correlation functions (\twoPCFs).

\subsection{Galaxy-Halo Connection} \label{sec:res_SMHF}
The \galacticus model assumes virial over-densities to define halo masses, but the measurements we want to compare to use $\Delta_c=200$, where $c$ refers to the critical over-density. Therefore we convert the halo masses \Mvir\ of our samples to the halo mass of our references \Mc\ following \citet[][Sec. 2.1]{Lokas&Mamon01}. Particularly, we use their Eq. (8) to calculate the ratio of the halo masses $P_{\Mhalo}=\Mc/\Mvir$ which depends on the halo concentration parameter \C\ as defined by Navarro-Frenk-White \citep[NFW,][]{Navarro97}. Since the \galacticus model does not provide this quantity nor the virial radius as outputs, we have to estimate the values using the fitting-formula of \citet[][Eq.(24)]{Klypin16_MD} and the corresponding values in Table 2 for $z=0.50$. We calculate the $P_{\Mhalo}$ for each galaxy separately, however the median over all ratios is $P_{\Mhalo} \sim 0.884 \pm 0.002$. Our estimated NFW concentration parameters can be found roughly in the range of $4 \lesssim \C \lesssim 6$ for $10^{13.3}<\Mc<10^{15.3}$ \Msun.\par

Note further that we refer to a ``central halo'' as the top-level dark matter halo in a certain merger tree and to ``central galaxies'' or ``centrals'' as the galaxies which reside in the centre of that haloes. From hereafter we exclude all orphan satellites because in the \galacticus model they are not connected to the current central halo anymore, but point to the dark matter halo they belonged to in the past (see \citealt[][A2]{Knebe17_MD} for clarification). Furthermore, their positions are not traced in the \galacticus model, but are assigned to the central galaxies they have been associated to previously. This introduces uncertainties when calculating correlation functions which we avoid by excluding them.\vspace{-0.3cm}

\begin{figure*}
    \includegraphics[width=12cm]{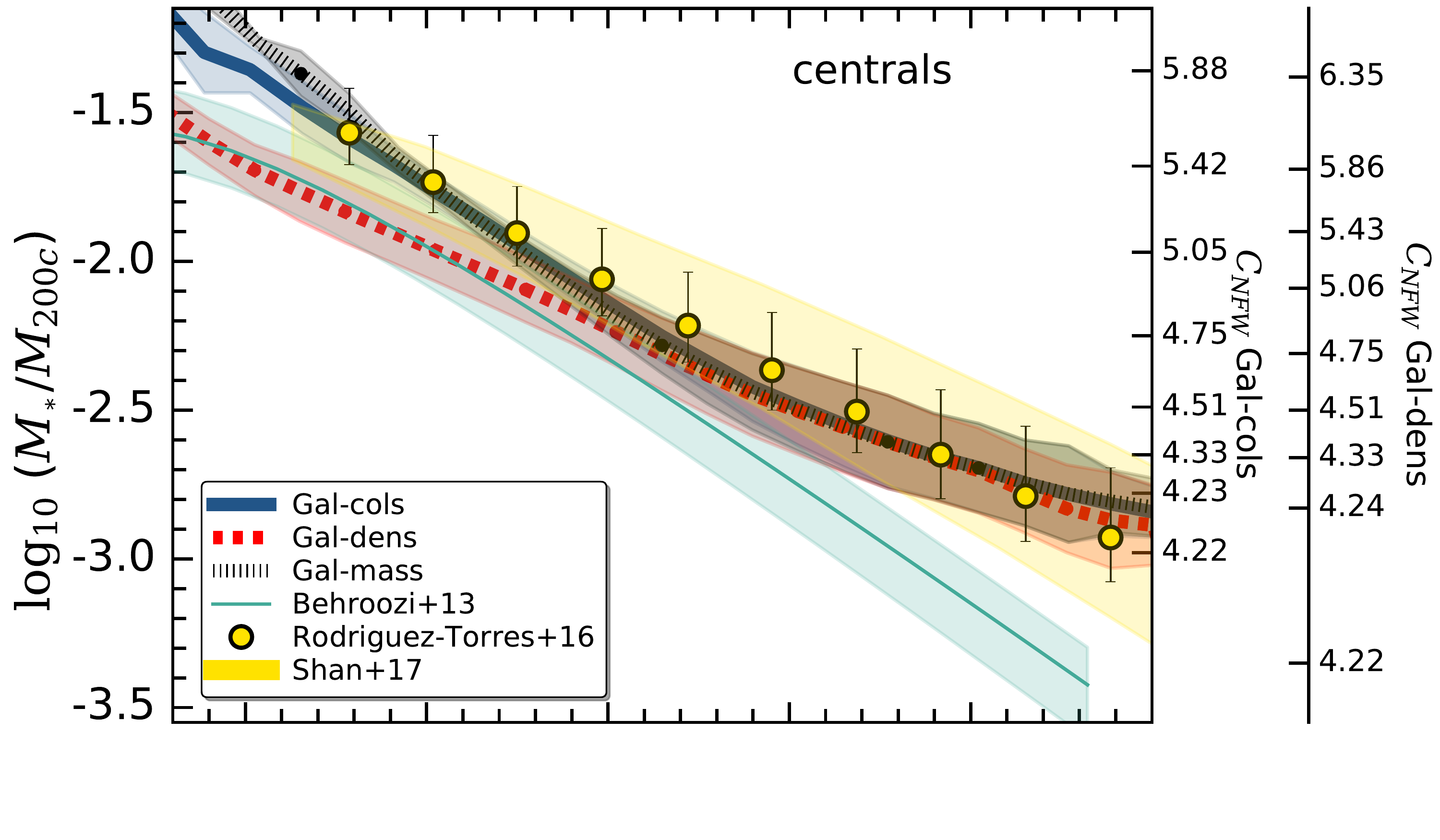}\vspace{-1cm}
    \includegraphics[width=12cm]{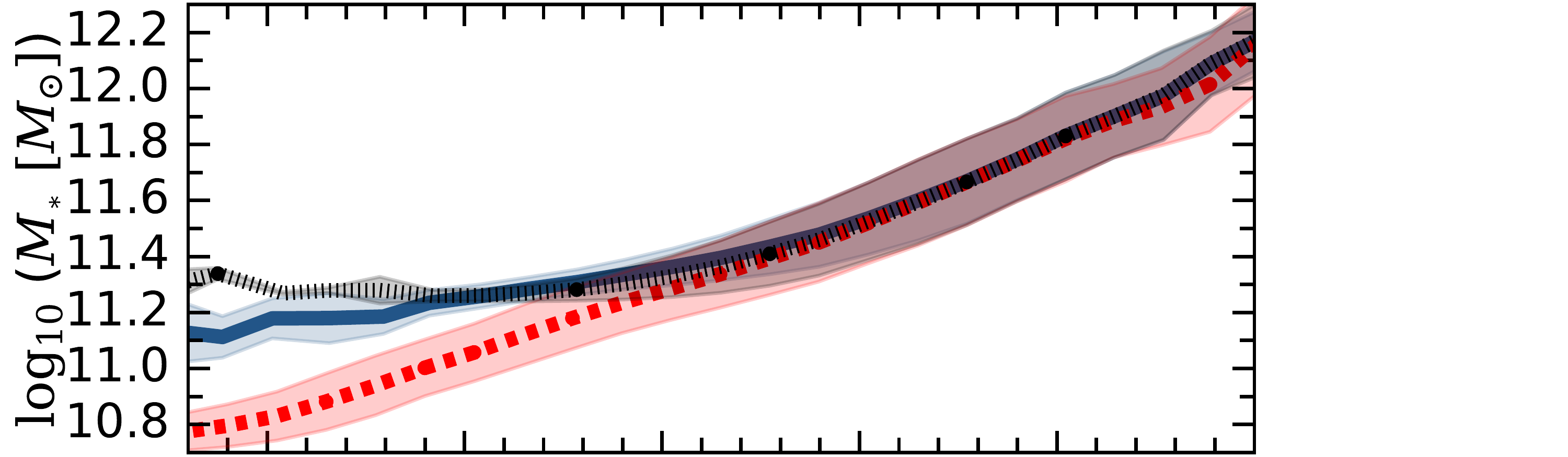}\vspace{-0.035cm}
    \includegraphics[width=12cm]{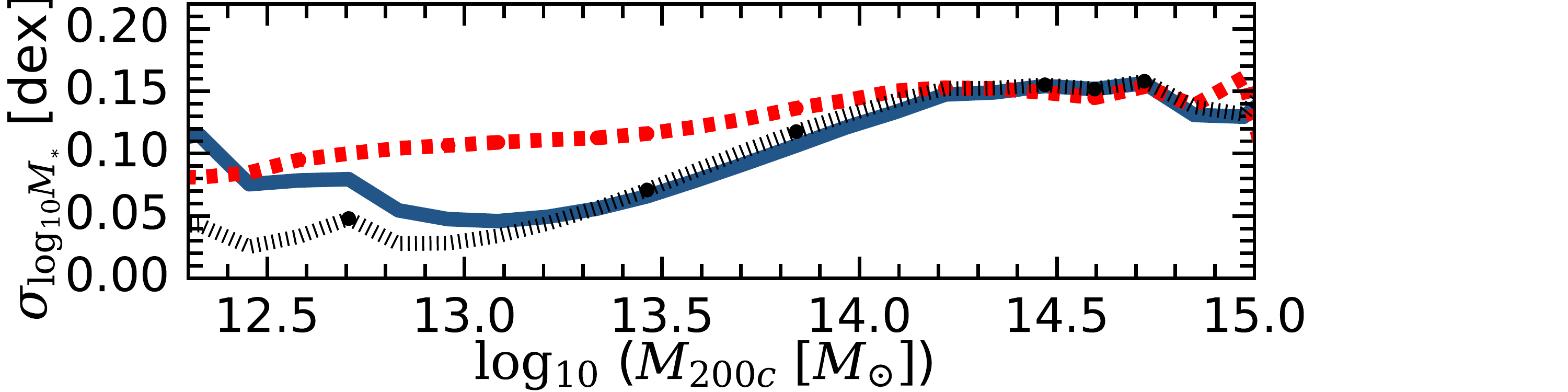}%
    \caption{\textit{Top}: \SHMF\ of central galaxies of \galacticus' CMASS-mock samples: \cm (blue solid line), \den\ (red dashed line), and \ma (grey dotted-dashed line). They are in excellent agreement with \blc with in $0.5<z<0.6$ (filled yellow circles), various HAMs realisations at $z \sim 0.55$ (shown as a thin green line) from \citet{Behroozi13c}, and weak-lensing observation from CFHT Stripe 82 ($0.4<z<0.6$) from \citet{Shan17} (shaded yellow area). The additional right y-axis represent the estimated halo concentration parameter \C\ for \cm and \den, respectively. \textit{Middle} and \textit{Bottom}: Stellar masses, \Mstar, and values for the intrinsic scatter, \sigM, respectively, as a function of \Mc\ for the same samples as shown in the \textit{top} panel.}\label{fig:SHMF}
    \vspace{-0.2cm}
\end{figure*}

\paragraph*{Stellar to halo mass ratio \Mstar/\Mc:}
In the \textit{upper} panel of \hyperref[fig:SHMF]{\Fig{fig:SHMF}}, we show \SHMF\ of our CMASS-mocks for central galaxies only (hereafter ``centrals'') compared to the halo abundance matching (HAM) model from \citet{Rodriguez-Torres16} based on the \textsc{BigMultiDark} simulation box with 2.5\hGpc\ side-length and clustering results from \boss-\cmass light cone (\blc, a mock light cone constructed with the sub-halo abundance matching modelling technique (sHAM) which reproduces \bossDR Large Scale Structure catalogue perfectly) within $0.5<z<0.6$. We further compare our SAM data to a compilation of various HAM realisations from \citet{Behroozi13c}\footnote{The data was modified to match the cosmology and initial mass function we assume in this paper.} at $z \sim 0.55$  and weak-lensing measurements from the Canada-France-Hawaii Telescope (CFHT) Stripe 82 from \citet{Shan17} within $0.4<z<0.6$, respectively. The additional y-axis on the right represents the estimated values for the NFW profile halo concentration, \C, for the two mock samples, \cm and \den, respectively. Note that we do not show an additional right axis for \ma because its values are similar to \cm. We report that our CMASS-mocks are in very good agreement with both, \blc and weak-lensing results e.g. \cm and \ma coincide with the data from the \blc to a high degree. However, \den\ agrees best with the HAM at low halo masses but then coincide with the other two samples at $\Mc\sim10^{13.5}$. In general we expect \galacticus' samples not to follow the HAM from \citet{Behroozi13c} because they use very different \SMF\ to build up their model (\texttt{PRIMUS} and \texttt{GALEX} \citealt{Moustakas13}\footnote{The difference between \texttt{GALEX} and \galacticus can be found in \citet[][Fig.1]{Knebe17_MD}.}). Their \SMF predicts less massive objects than those from \boss as we also found for \denS. \par 

Additionally, we tested the impact on the results using \galacticus native definition of over-densities ($\Delta_{\rm BN98}$) and their corresponding halo mass \MBN. The impact on the \SHMF\ is small but visible on most massive haloes, but within the error estimations.\vspace{-0.3cm}

\paragraph*{Star formation efficiency:}
In the \textit{middle} panel of \hyperref[fig:SHMF]{\Fig{fig:SHMF}} we plot the corresponding \Mstar\ at fixed halo mass and show that the stellar masses truly stays constant for increasing halo masses up to $\Mhalo \sim 10^{13.5}~\Msun$ considering \cm and \ma. Then \Mstar\ increases continuously which explains the shallower slope of the \SHMF\ in the high-mass regime. That means that the most massive haloes in the CMASS-mocks host galaxies which have been producing stars more efficiently in their lifetime compared to the \blc or the HAM.\vspace{-0.3cm}

\paragraph*{Intrinsic scatter \sigM:}
In the \textit{lower} panel of \hyperref[fig:SHMF]{\Fig{fig:SHMF}} we plot the intrinsic scatter between stellar and halo mass, \sigM, for \galacticus CMASS-mock samples. As reported in the literature \citep[e.g.][]{Moster10,More11_sats3,Leauthaud11b,Tinker17}, the relation between the stellar and halo mass is not one-to-one, meaning that the most massive haloes do not host the most massive galaxies (as requested by e.g. HAM models). Furthermore, two haloes with the same mass can host different galaxies with different stellar masses due to distinct assembly history, environmental effects, or feedback mechanisms (to name only a few). The distribution in stellar mass at fixed halo mass is called ``intrinsic (log-normal) scatter'' and is given by the standard deviation of logarithmic base 10 stellar mass at that halo mass \citep{Tinker13}. As shown in the \textit{lower} panel of \hyperref[fig:SHMF]{\Fig{fig:SHMF}}, \sigM\ varies from sample to sample. It depends strongly on halo mass for \cm and \ma and drops to a minimum at $\Mc\sim 10^{13}$ \Msun. This means that for growing halo mass, the stellar mass of galaxies residing in these haloes stays constant until the halo reaches a certain mass threshold. \den\ does not exhibit such a threshold or minimum, but shows an almost constant scatter of $\sigM\sim0.15$ dex for haloes with masses of $\Mc>10^{14}$ \Msun\ and then declines smoothly to $\sigM=0.09$ dex for $\Mc<10^{14}$ \Msun. Due to the down-sampling process on the \SMF\ of \boss, \den\ exhibits a higher fraction of low-mass haloes than the other CMASS-mocks which is reflected in the intrinsic scatter.\vspace{-0.3cm}

\paragraph*{Halo occupation distribution (HOD):}
\begin{figure*}
    \hspace{-0.55cm}
    \includegraphics[width=6.8cm]{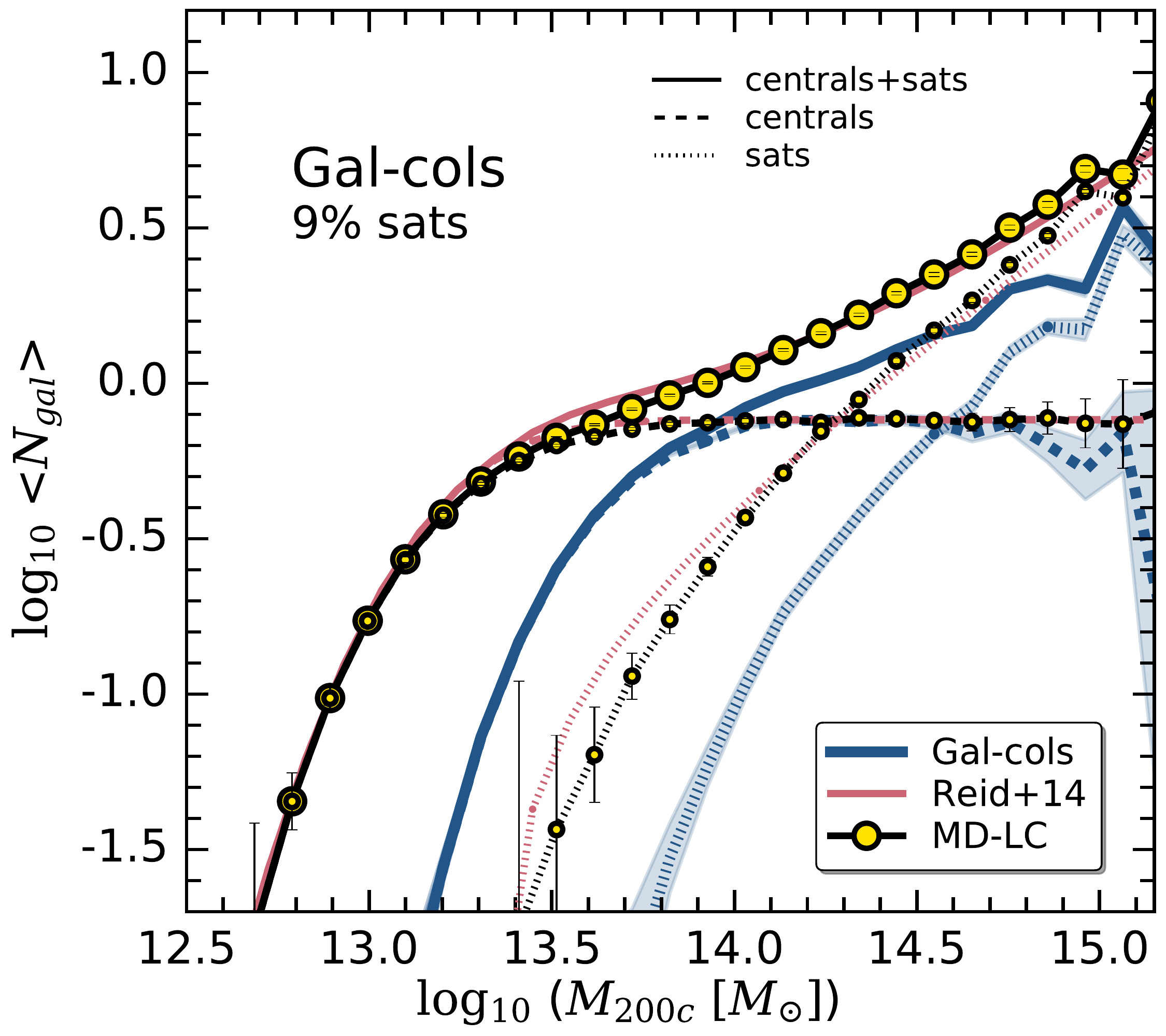}%
    \hspace{-1.18cm}
    \includegraphics[width=6.8cm]{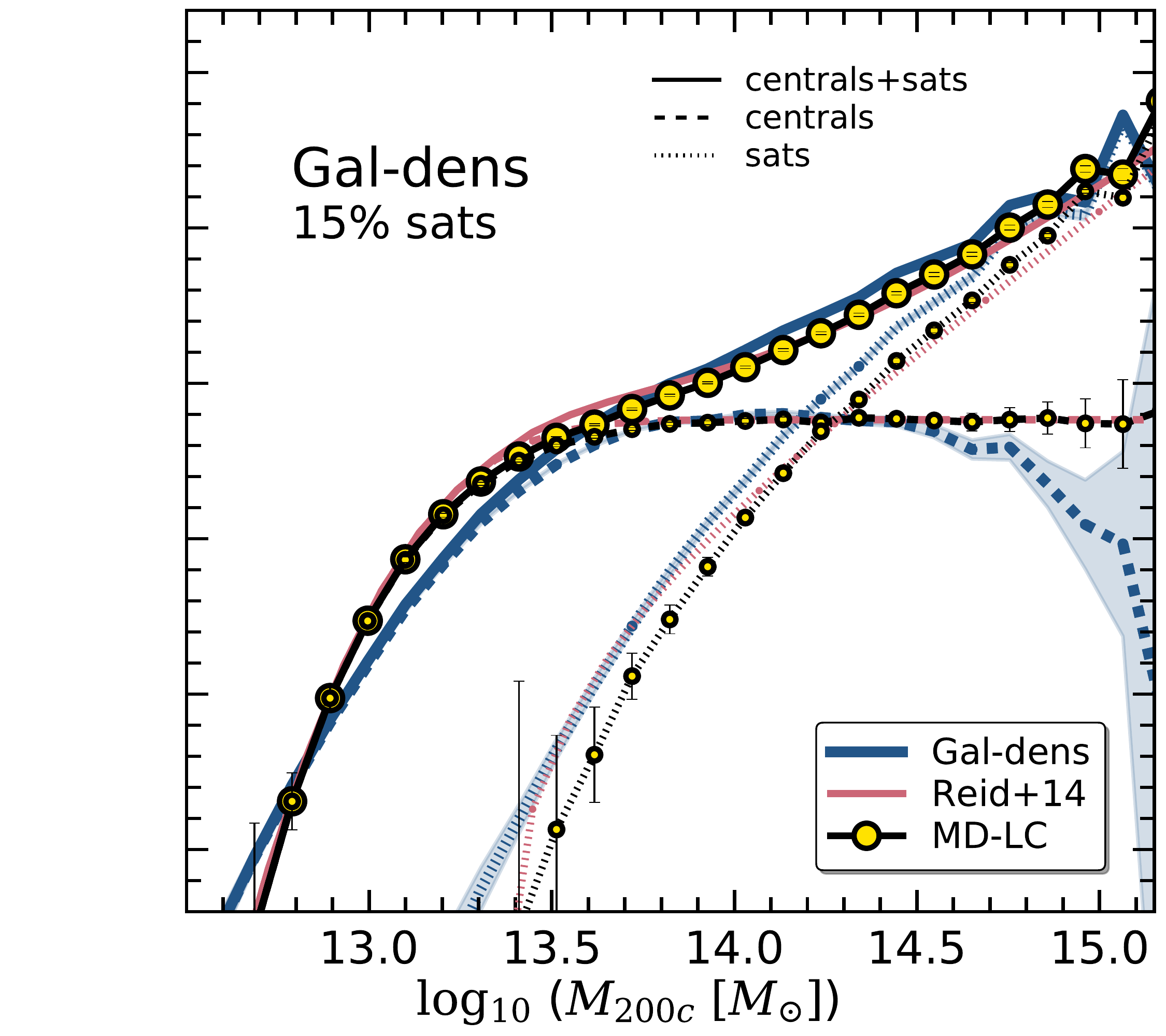}%
    \hspace{-1.18cm}
    \includegraphics[width=6.8cm]{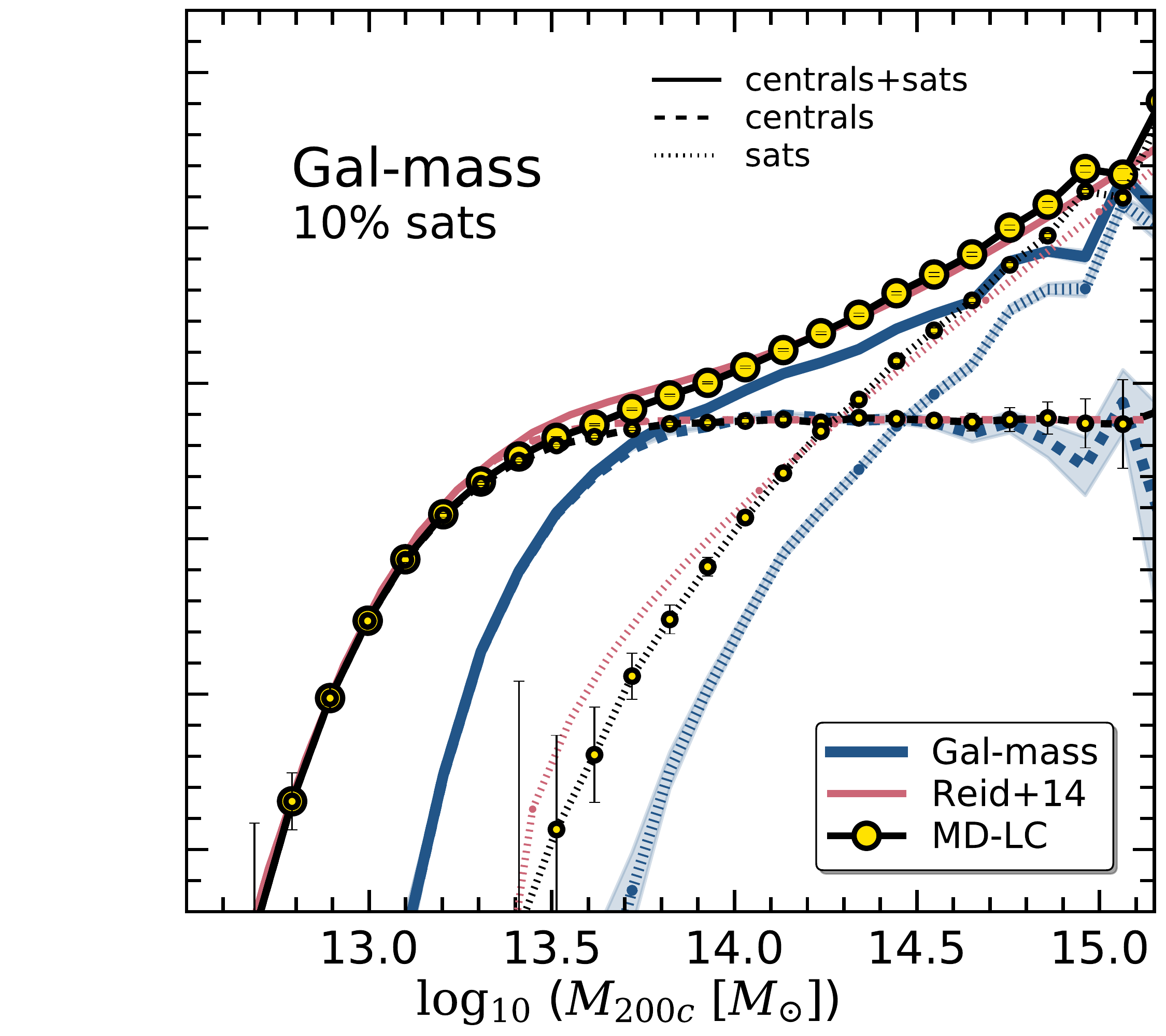}\vspace{-0.2cm}%
    \caption{Halo occupation distributions split into their components where solid lines represent centrals+satellite galaxies (short: centrals+sats), long dashed lines represent centrals and short dashed lines satellites only. \galacticus' samples are shown as thick blue lines in the panels from left to right: \cm, \den, and \ma. We compare to the HOD model from \citet{Reid14} (thin red lines) and to the \blc based on abundance matching from \citet{Rodriguez-Torres16} (yellow filled circles on black thin lines) using the same line style keys as \galacticus for their HOD components. \textless \Ngal \textgreater\ is the mean number of galaxies of a halo with a certain mass \Mc.}\label{fig:HOD}
    \vspace{-0.4cm}
\end{figure*}

As a second tool to describe galaxy-halo connection, we present the HOD, the mean number of galaxies per halo, \textless\Ngal\textgreater, as a function of the halo mass, \Mc. The contribution to the form of the HOD can be divided into central galaxies, modelled as a step function, and satellites, following a power law \citep{Berlind03_HOD,Zheng05_HOD}. In \hyperref[fig:HOD]{\Fig{fig:HOD}} we show in three panels the HOD components for our CMASS-mocks from left to right: \cm, \den, and \ma.\par

Furthermore, we compare to a HOD-fit from $N$-body simulations constructed from \texttt{SDSS-III DR10} data \citet[][their \textsc{MedRes0} simulation box]{Reid14} modified to the number density of \cmass at $z=0.56$ (by applying a factor of $1/1.31$ to their HOD in order to correct from their adopted number density to $n=1.02 \times 10^{-4}$ \MpcV). We use their best-fitting model from an adaptation of \citet{Zheng05_HOD}. We further compare to the first \textsc{MDPL} cosmological simulation. This simulation uses the same cosmology and parameters as \MDPL, like 1\hGpc\ side-length of the box and we constructed the HODs by applying the same HAM-recipe as described in \citet{Rodriguez-Torres16} for the \bmd.\par 

All \galacticus CMASS-mock samples show highly diverse shapes of their HODs where the \den\ follows our adopted references best. In the high-mass end and for the contribution of satellites, \den\ agrees with the observations better than the other two. Although \cm and \ma show abundances of satellites in agreement with observations ($\sim10$\%, see \hyperref[tab:tarsel]{\Tab{tab:tarsel}}), \den\ is with 15\% satellites the only sample where the HOD of satellites is comparable to the data.\par

The ``knees''\footnote{The probability that half of the haloes host at least one galaxy, equal to \Mmin.} of the HOD differ a lot between the CMASS-mock sample being estimated by eyeballing: $\Mhalo \sim 10^{13.7}$ \Msun\ for \cm and $\Mhalo \sim 10^{13.5}$ \Msun\ for \den\ and \ma, respectively, and to the observation with $\Mmin=10^{13.180}$ \Msun. The transition between a halo hosting zero to at least one galaxy is more gradually for \den\ and more steep for \cm and \ma. The halo mass where a halo cannot host at least one satellite anymore (see short dashed line) varies from $\Mhalo \sim 10^{13.3}$ \Msun\ (\den) to $\Mhalo \sim 10^{13.8}$ \Msun\ (\cm) and corresponds to $\Mcut=10^{13.328}$ \Msun\ for the observations and \blc, respectively.\par

All CMASS-mock samples show a similar \Mc\ $M_1$ \footnote{The probability to find 1 satellite/halo drops to $<1$ (equal to $M_1$).} between $10^{14.3}<\Mc<10^{14.7}$ \Msun\ compared to the data with $\Mc\sim10^{14.2}$ \Msun. A large plateau also corresponds to large $M_1/M_{min}$-ratio being $\sim10$ for \cm and \ma and $\sim6$ for \den, compared to our references with $\sim11$. This ratio has a significant impact on the shape of the correlation function \citep{Benson00} meaning that galaxies within a wide range of mass or luminosity exhibit a power-law correlation functions \citep{Zheng05_HOD}.\par

The HODs for centrals (blue thick dashed lines) show incompleteness at the highest halo mass for all \galacticus CMASS-mocks, mainly due to the limited volume of the simulation box. We also see that the \den\ CMASS-mocks lacks significantly in high-mass central galaxies which have been excluded during the down-sampling procedure. However, the abundance of the satellites are in complete agreement with our references. Furthermore, the fact that \cm and \ma show a smaller scatter in stellar mass than \den\ can be directly read from the HODs of the satellites.

\subsection{2-point Correlation Function (\twoPCF)}\label{sec:res_2pf}
\begin{figure*}
    \centering
    \includegraphics[width=9.3cm]{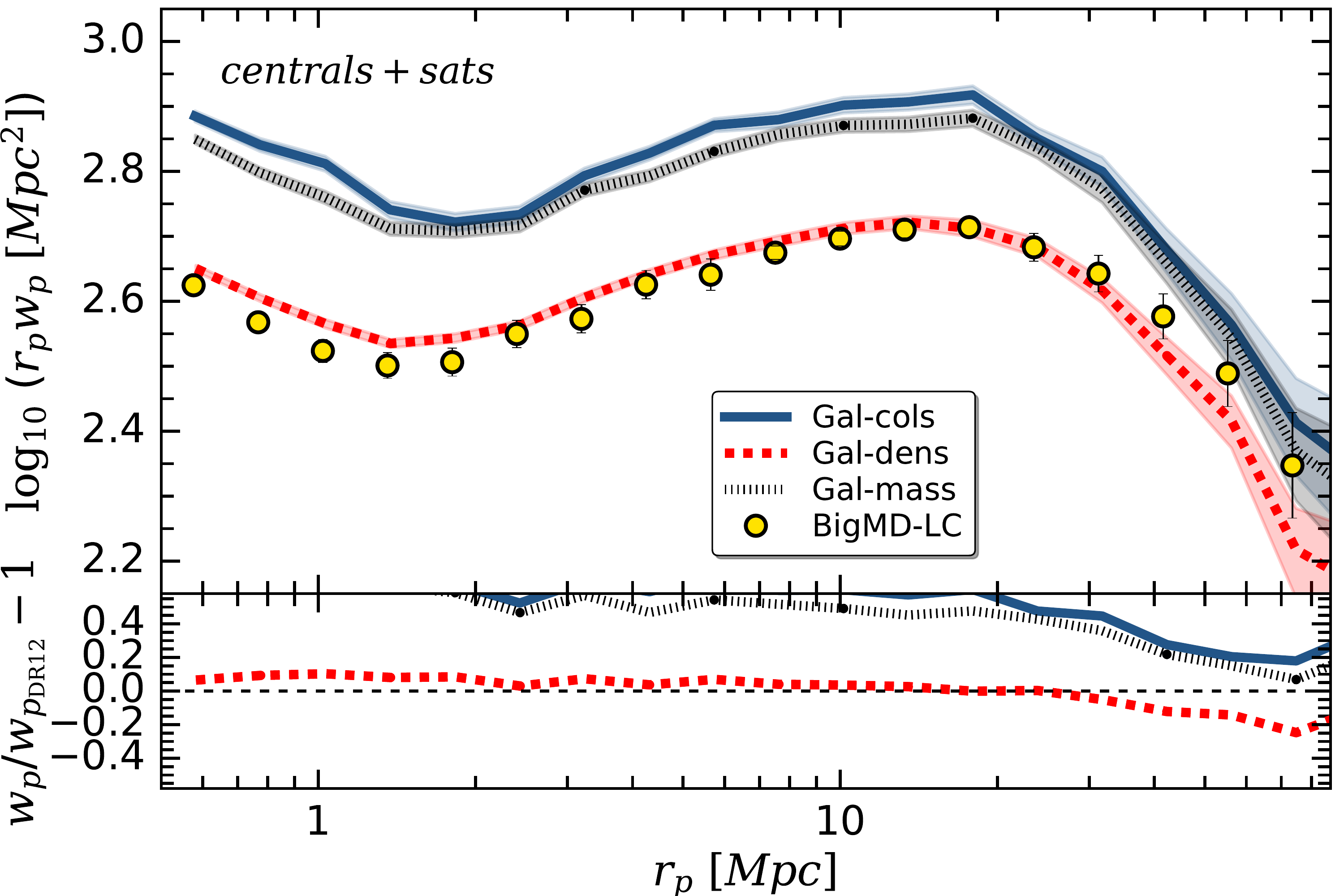}%
    \hspace{-1.25cm}
    \includegraphics[width=9.3cm]{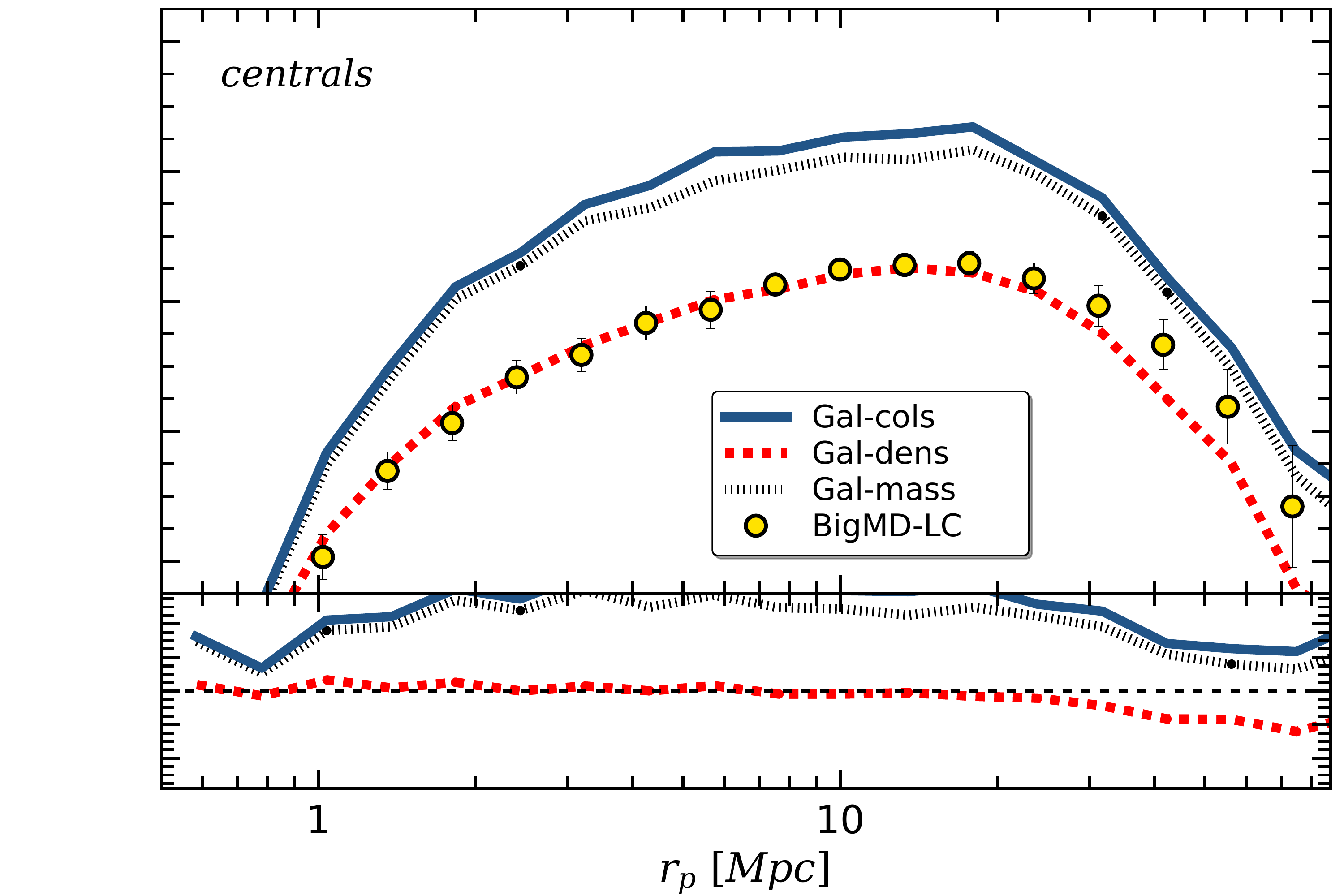}\vspace{-0.2cm}%
    \caption{The projected 2-point correlation function for \galacticus CMASS-mock samples: \cm (blue solid line), \den\ (red dashed line), and \ma (grey dotted-dashed line) at redshift $z=0.56$ compared to the \blc (filled yellow circles) for centrals+sats (left) and centrals only (right). The amplitude and shape of the \twoPCF is highly diverse for our different CMASS-mock samples and also depends on the galaxy type. The best reproduction of the observations was achieved in general by the \cmS.}\label{fig:wp}
    \vspace{-0.4cm}
\end{figure*}

In this section we present our results for the projected 2-point correlation function (\twoPCF) for our CMASS-mock samples. We use the \textsc{corrfunc} software package\footnote{\url{http://corrfunc.readthedocs.io/en/master/index.html}\label{ftn:corrfunc}} from \citet{corrfunc} and the standard \citet{Landy&Szalay93} estimator to calculate the functions. We produce \twoPCFs with 20 log-spaced bins in the range of $0.5<\rp<150$ \Mpc\ with an integration length of $\pim=150$ \Mpc. We also show the influence of the galaxy type by calculating correlation functions for central and satellite galaxies (short: centrals+sats) and centrals only. \vspace{-0.3cm}

\paragraph*{\twoPCFs for different galaxy types:}
In \hyperref[fig:wp]{\Fig{fig:wp}} we present \twoPCFs for centrals and satellite galaxies (\textit{left}) and centrals only (\textit{right}). We compare to the \blc\footnote{Note that we do not compare directly with observations because \citet{Rodriguez-Torres16} already showed that the \blc agrees very well with \boss (see their Fig. 10). Therefore we treat \blc data like observations in this work. Furthermore, we calculated the \blc data points using a rescaled light cone to match the box size of \MD.} within $0.5<z<0.6$, using the same data and treatment as described in \citet[Sec 5.1.]{Rodriguez-Torres16}. We estimate the uncertainties of our CMASS-mocks for centrals and satellites using 200 realisations of the \textsc{MD-Patchy} mocks \citet{Kitaura16}. In order to account for the smaller box size-length of \MDPL\ we used the \textsc{MD-Patchy} mocks down-scaled to 1\hGpc. We note that, we did not construct error bars for centrals only because the \textit{MD-Patchy} code does not distinguish between central and satellites. \par

In the \textit{lower} panel of \hyperref[fig:wp]{\Fig{fig:wp}}, we show the residuals for \galacticus CMASS-mock samples compared to the \blc. The CMASS-mocks \cm and \ma fail to reproduce the \twoPCF of the \blc, independently if considering centrals and satellite galaxies together or centrals only.  However, the shape of their functions are similar but they exhibit a constant shift of $\sim0.5$ dex towards higher amplitudes compared to \blc. Only \den\ is in very good agreement with the data over a large range of \rp\ for both, centrals and satellites and centrals only. \par

If we include low-mass objects as in the \denS\ the clustering amplitude is reduced at all scales except of the largest with $\rp>40~\rm Mpc$ in full agreement with the results of the the HODs in \hyperref[fig:HOD]{\Fig{fig:HOD}}. The \textit{left} panel of that figure shows that low-mass halos are underrepresented in the \cm' HODs resulting in a higher amplitude of the correlation functions in \hyperref[fig:wp]{\Fig{fig:wp}}, because only the distances between the most massive objects have been taken into account. \den' HOD (\textit{middle} panel) and \twoPCF\ agree well with both, \mlc\ in \hyperref[fig:HOD]{\Fig{fig:HOD}} and \blc in \hyperref[fig:wp]{\Fig{fig:wp}}, because more low-mass objects could enter the sample. This is true for centrals and satellite galaxies or for centrals only. We therefore investigate which galaxies contribute the most to the correlation function by selecting subsamples for different subsequent stellar mass cuts. We further hereafter drop the discussion of the \maS because the results from it is almost identical to that from the \cmS.\vspace{-0.3cm}

\paragraph*{\twoPCFs of various subsequent \Mstar\ cuts:}
\begin{figure} 
    \includegraphics[width=8.4cm]{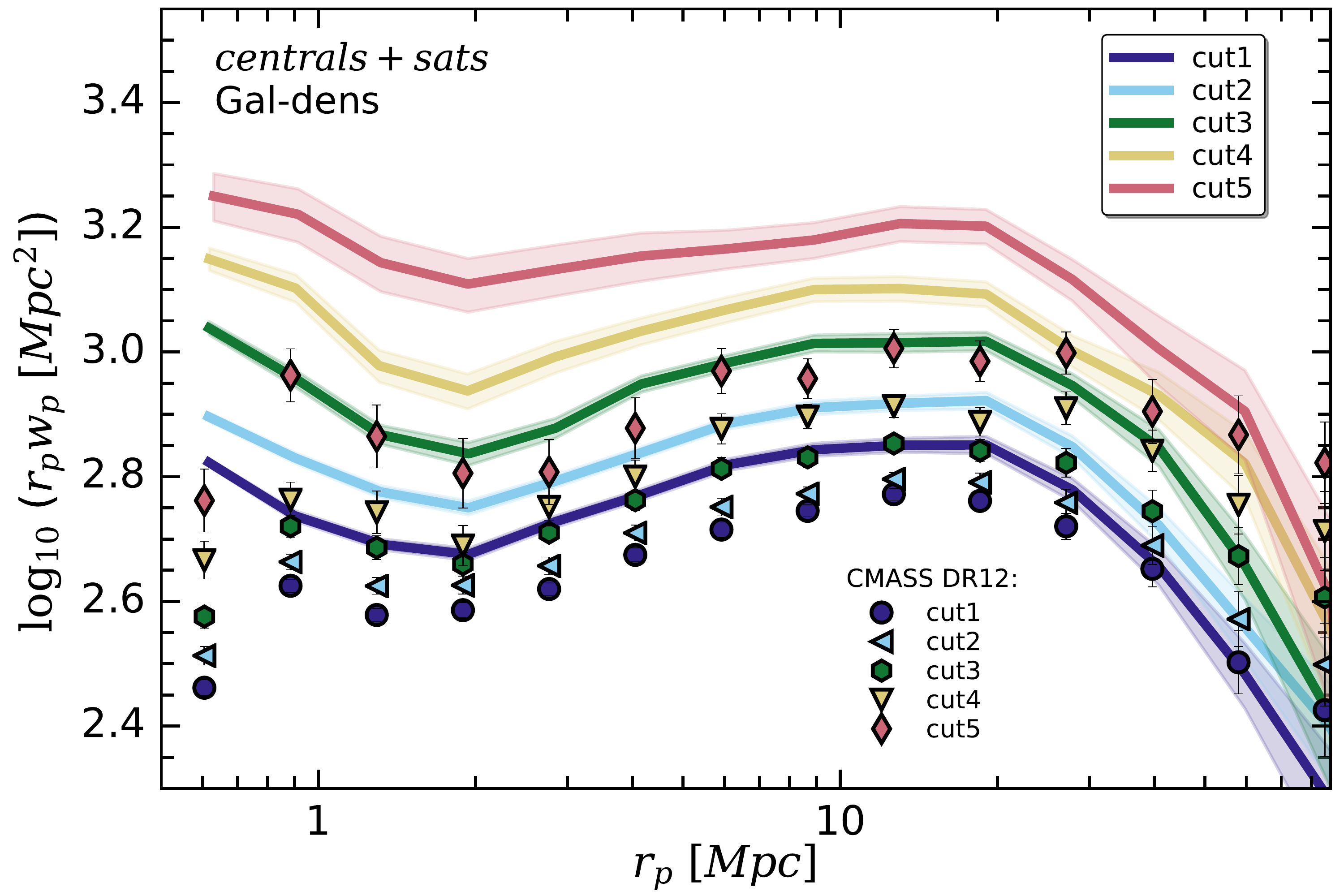}\vspace{-0.2cm}%
    \caption{Projected 2-point correlation functions of sub-samples of \galacticus' CMASS-mock \den\ (solid lines) using the subsequent \Mstar\ cuts (indicated by the keys) compared to \bossDR (markers).}\label{fig:wp_mass}
    \vspace{-0.4cm}
\end{figure}

We show the \twoPCFs of sub-samples of the CMASS-mock sample \den\ in \hyperref[fig:wp_mass]{\Fig{fig:wp_mass}}. The sub-samples were constructed by the applying a subsequent stellar masses cuts in \logT (\Mstar\ $[\Msun]$): (cut1) 11.21, (cut2) 11.31, (cut3) 11.41, (cut4) 11.51, and (cut5)  11.61. We use again 200 realisations of the \textsc{MD-Patchy} mocks for the estimation of the uncertainties as in \hyperref[fig:wp]{\Fig{fig:wp}}. Note that we only present results for \den\ because only this sample provides a sufficient number density of galaxies. We can see in the figure that, modelled and observed galaxies are in poor agreement with each other. In order to improve the clustering we tried to fix the number density $n$ of \galacticus' sub-samples in order to match those of \bossDR. This experiment only improved the \twoPCF slightly. 

\section{Discussion} \label{sec:discussion}
Before we discuss our results we want to add a few notes about the influence of \galacticus native tuning and model configuration. Most importantly, \galacticus has not been specifically calibrated on \MDPL, but its most favorables parameter set and configuration were used. Although \galacticus was tuned to match the $K,b_j$-band luminosity functions at $z=0$ and the local colour-magnitude diagram at $z=0.1$, its luminosities and colours do not perfectly match the \cmass galaxy properties. Therefore, we examine if alternative approaches to select a CMASS-mock (e.g. a cut in stellar mass) would be a convenient approach to bypass this problem. In general the \cm and \ma samples agree very well with each other (see \hyperref[fig:dis_fidplots]{\Fig{fig:dis_fidplots}} or results of \SMF, \SHMF, HOD, or \twoPCF), but both exhibit too low number densities compared to \cmass and do not reproduce the \twoPCF as shown in \hyperref[fig:wp]{\Fig{fig:wp}}.\par

\textit{Why does a density-selected sample work better?} Firstly, \den\ exhibits by construction the same number density as \cmass. Secondly, although  \den' galaxies are 1.5-2 magnitudes fainter in the \iband than \cm, \ma, and \cmass (as shown in the \textit{upper} panel of \hyperref[fig:dis_fidplots]{\Fig{fig:dis_fidplots}}), their stellar masses are fully comparable\footnote{\den\ is located within the 95\% confidence level contour of \cmass in the \gi colour plane as shown in the \textit{lower} panel of \hyperref[fig:dis_fidplots]{\Fig{fig:dis_fidplots}}.} and should have satisfied the CMASS colour-magnitude selection criteria, but due to their lower brightness they did not enter the sample selection.\par

\begin{figure}
    \begin{center}
    \includegraphics[width=8.4cm]{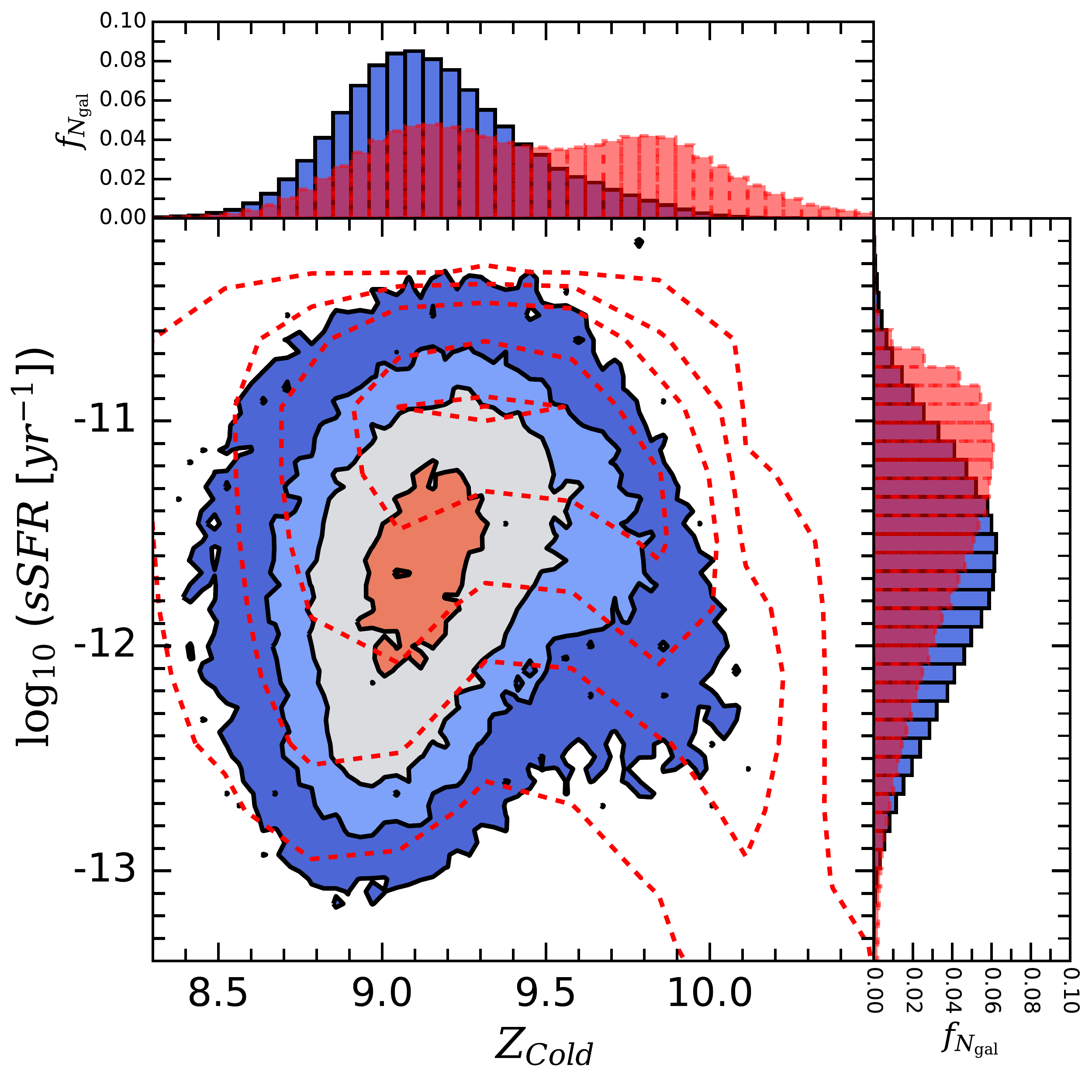}\vspace{-0.2cm}%
      \caption{Relation between gas-phase metallicity $\Zcold$ and \sSFR\ for central galaxies of the \cmS (filled coloured contours) and the \denS\ (red dashed contours) at $z=0.56$.}\label{fig:ssfr2zcold}
      \vspace{-0.4cm}
    \end{center}
\end{figure}

\textit{What are the properties of \den\ galaxies?} We can divide \den\ into two distinct populations (A) and (B) using a sliding cut in \SFR\ depending on \sSFR\footnote{The following conditional equation divides the sample into Population (A) and (B):\vspace{-0.3cm}
\begin{equation}
	\logT (\SFR~[\Msunyr])
	\begin{cases}
	<\delta & \text{Pop (A)} \\
	>\delta & \text{Pop (B)} 
	\end{cases}
	\text{where } \delta= \frac{\logT~(\sSFR [\rm yr^{-1}])+11.16}{1.12}
\end{equation}\vspace{-0.3cm}}. %
Population (A)  galaxies are low-starforming ($\SFR\sim0.05$ \Msunyr) and live in low-mass haloes ($\Mc<10^{13.3}$ \Msun) while Population (B) are starforming ($0.1<\SFR<0.3$ \Msunyr) residing in most massive haloes ($\Mc>10^{13.3}$ \Msun). We find a strong dependency on halo mass at fixed \sSFR\ where low-mass haloes have a linear relation between \SFR\ and \sSFR, while the high-mass haloes exhibit larger \SFRs\ at fixed \sSFR. Furthermore, certain galaxy properties related to star formation can be clearly mapped onto Population (A) or (B) but other properties such as \Mstar\ or \ri-colour are continuously distributed. This trend is particularly interesting because it shows the importance of secondary parameters related to the clustering besides halo mass as suggested by \citet{Wang13}. \par 

\textit{How do gas-phase properties divide the sample into two distinct populations?} In \hyperref[fig:ssfr2zcold]{\Fig{fig:ssfr2zcold}} we show the gas-phase metallicity \Zcold\footnote{$\Zcold=8.69+\log_{10}(M_{Z, \rm Cold}/\Mcold)-\log_{10}(Z_{\odot})$, where $M_{Z,\rm Cold}$ is the mass of metals in the cold gas-phase. \Zcold\ is normalised by the metallicity of the Sun $Z_{\odot}=0.0134$ \citep{Asplund09}, while the factor $8.69$ \citep{Allende01} corresponds to its oxygen abundance.}, a proxy for gas-cooling and star formation \citep{Lebouteiller13}, for central galaxies. The two populations (A) and (B) are reflected in the bimodal distribution of \Zcold\ where $\sim80\%$ of Population (A) shows $\Zcold>9.5$ and only 20\% lower values with $\Zcold\sim9.5$. The opposite is true for Population (B). Common studies of fundamental relations between metallicity, mass, and star formation suggest that less/more massive galaxies have also lower/higher metal abundances \citep{Lara-Lopez10a,Yates12}. Our results show that Population (B)'s galaxies are more massive but have lower metal abundances. This ``turnover'' was also reported by \citet{Yates12} for modelled galaxies at $z=0$ and is possibly linked to the infall of metal-poor gas after a gas-rich merger. In \citet{Yates&Kauffmann14} the same authors studied massive galaxies and divide them into an ``enriching'' and a ``diluting'' sample, the later show similar trends as our Population (B): low \sSFR, lower \Zcold\, and higher $M_{\rm BH}$. Furthermore, our findings are in total agreement with \citet{Lara-Lopez13} showing that galaxies with low \sSFR\ have high/low values of \Zcold\ when \Mcold\ is low/high. We emphasise that the distinct separation of the two populations could give clues about galaxy evolution in the context of the origin of the fundamental luminosity/mass-metallicity relation, merger-induced star formation, or ``downsizing'' \citep[][see their Sec.1 for comparison]{Mannucci10}.\par

\textit{How do the Populations (A) and (B) relate to environment?} We expect that Population (A) fixes the clustering amplitude due to their environment as well as their number density. To this extent we apply the {\sc Vweb} method (see \hyperref[app:A]{\App{app:A}} for details) to the underlying dark matter \MDPL\ simulation.  We show in the second column of \hyperref[tab:env_props]{\Tab{tab:env_props}} that more galaxies in the \cm\ sample (61 \%) are assigned into knots than in the \denS\ (52 \%). We detect a clear environmental dependency of this sample where Populations (A) is dominated by filament galaxies (62\% with only 26 \% in knots), while Population (B) has more galaxies in the knots (54\%) than in the filaments (44\%). 

\begin{table*}
 \begin{center}
  \setlength{\tabcolsep}{3.1pt}
   \begin{tabular}{llccccccc}
	\hline
	sample name \& & environment & fraction & \logT(\Mc\ [\Msun]) & \logT(\Mstar\ [\Msun]) & \logT(\sSFR\ [$\rm yr^{-1}$]) 
	& \Zcold & \logT($\Mcold/\Mstar$) & \logT(\Mbh\ [\Msun])\\
	population & & of galaxies & & & & & & \\
	\hline
	 \hline	  
	\cm		& knot & 0.61	   & 13.79$^{+0.21}_{-0.19}$ &  11.44$^{+0.11}_{-0.09}$ & -11.77$^{+0.35}_{-0.37}$ & 9.10$^{+0.18}_{-0.16}$ & -1.14$^{+0.20}_{-0.26}$ &  8.65$^{+0.21}_{-0.19}$\\
	\cm		& filament & 0.37 & 13.57$^{+0.19}_{-0.17}$ &  11.37$^{+0.09}_{-0.05}$ & -11.52$^{+0.30}_{-0.34}$ & 9.18$^{+0.21}_{-0.17}$ & -1.25$^{+0.22}_{-0.30}$ &  8.53$^{+0.19}_{-0.16}$\\
	\hline
	\den		& knot & 0.52	   & 13.55$^{+0.27}_{-0.28}$ &  11.28$^{+0.15}_{-0.14}$ & -11.53$^{+0.36}_{-0.42}$ & 9.25$^{+0.36}_{-0.23}$ & -1.35$^{+0.31}_{-0.65}$ &  8.43$^{+0.26}_{-0.26}$\\
	\den		& filament & 0.41 & 13.22$^{+0.23}_{-0.26}$ &  11.14$^{+0.13}_{-0.14}$ & -11.36$^{+0.32}_{-0.44}$ & 9.55$^{+0.30}_{-0.35}$ & -1.83$^{+0.54}_{-0.77}$ &  8.20$^{+0.25}_{-0.24}$\\
	\den\ 	Pop (A)	& knot	& 0.26   & 13.13$^{+0.32}_{-0.34}$ &  11.05$^{+0.07}_{-0.09}$ & -11.40$^{+0.38}_{-0.53}$ & 9.76$^{+0.19}_{-0.29}$ & -2.32$^{+0.61}_{-0.71}$ &  8.05$^{+0.19}_{-0.17}$\\
	\den\	Pop (A)	& filament & 0.62 & 13.01$^{+0.22}_{-0.22}$ &  11.02$^{+0.08}_{-0.11}$ & -11.40$^{+0.39}_{-0.58}$ & 9.80$^{+0.19}_{-0.25}$ & -2.42$^{+0.57}_{-0.74}$ &  8.00$^{+0.17}_{-0.16}$\\
	\den\	Pop (B)	& knot	& 0.54   & 13.69$^{+0.23}_{-0.21}$ &  11.36$^{+0.13}_{-0.09}$ & -11.58$^{+0.35}_{-0.40}$ & 9.13$^{+0.21}_{-0.17}$ & -1.18$^{+0.21}_{-0.31}$ &  8.56$^{+0.22}_{-0.20}$\\
	\den\	Pop (B)	& filament & 0.44 & 13.46$^{+0.18}_{-0.16}$ &  11.29$^{+0.09}_{-0.07}$ & -11.33$^{+0.26}_{-0.32}$ & 9.23$^{+0.23}_{-0.19}$ & -1.33$^{+0.26}_{-0.34}$ &  8.42$^{+0.20}_{-0.16}$\\
	 \hline	 
	 \hline	 
	(i)  & (ii) & (iii) & (iv) & (v) & (vi) & (vii) & (viii) & (ix) \\
	\hline
    \end{tabular}
  \end{center}
  \caption{The table summarises the median and 1$^{st}$ (sub-scripted) and 3$^{rd}$ (super-scripted) quartile values of various galaxy properties for central galaxies in different environments and mock galaxy samples. Column (i) states the name of the CMASS-mock sample (and population if given). Thereby ``Pop (A)'' refers to Population (A) and ``Pop (B)'' to Population (B). Column (ii) indicates the environment (knot or filament) and (iii) their corresponding fraction. Results for the median values of halo mass \Mc, stellar mass \Mstar, specific star formation rate \sSFR, gas-phase metallicity \Zcold, cold-gas fraction $\Mcold/\Mstar$, and black hole mass \Mbh, respectively, are given in columns (iv)-(ix). Note that we only analysed galaxies in knots and filaments if their number of objects is significantly high, otherwise results for the whole sample is give as for \cm.}
  \vspace{-0.4cm}\label{tab:env_props}
\end{table*}

\textit{Do galaxy properties have a dependency on environment?} Besides the number fraction of galaxies, we further detail the sample properties in different environments in \hyperref[tab:env_props]{\Tab{tab:env_props}}. Galaxies in filaments generally tend to have lower halo, stellar and black hole masses as well as \sSFR\ and cold-gas fraction compared to the ones (from the same sample) in knots, while the cold-gas metallicity is normally higher in filaments than in knots. It is worth to note that Population (A) has significantly smaller halo mass, stellar mass, cold-gas fraction, and black hole mass than Population (B) in both environments, but significantly higher cold-gas metallicity in (A) than in (B).\par 

\textit{What conclusion can we draw from the environmental dependency of galaxy properties?} The star formation is not sufficiently suppressed in Population (B) and the most massive galaxies which should be ``red-and-dead'' are still starforming at a low rate. Therefore, \galacticus\ shows a higher abundance in the high-mass end of the \SMF\ compared to the observed \cmass galaxy sample. Furthermore, most of the low-\SFR\ galaxies in the \den\ sample live in the filaments in Population (A) with relatively lower \Mbh\ and \Mcold. They are located in haloes with suppressed star formation and could not grow in mass enough to exhibit brighter luminosities. This scenario is supported by the fact that Population (A) of \den\ has small contents of cold gas and as smaller cold-gas fractions in both knots and filaments, compared to Population (B). We cannot explicitly say why \Mcold\ is significantly smaller but it would imply that the quenching process in \galacticus\ is mostly dominated by tidal stripping of the cold gas instead of AGN feedback. We find it further interesting that half of the galaxies of this population exhibit higher gas-phase metallicities. We could speculate that the two populations (A) and (B) might have formed at different times and evolved differently due to their environment (see ``environmental quenching'' of star formation e.g. \citealt[][]{Tomczak18}) or halo masses (see ``halo quenching'' of low-mass central galaxies e.g. \citealt{Tal14}). Different evolutionary paths (as \citealt{Montero-Dorta17_assembly} have shown for \boss) might have contributed to the variations in the intrinsic scatter and could also provide a signal of the assembly bias, however, further studies are required to provide proof of that hypothesis. \par

\textit{How is the environmental dependency reflected in the clustering?} We expect that the different quenching processes have a crucial impact on the intrinsic scatter in stellar mass at fixed halo mass, \sigM, which in return has an impact on the clustering amplitude. Compared to other works we report that the values of the intrinsic scatter of \cm and \ma with 0.1 dex and \den\ with 0.15 dex depending on the halo mass. Those results are similar to \citet{Rodriguez-Torres16}, who found a scatter of 0.14 dex for their CMASS abundance matching \blc. However, \citet{Shankar14} stated that an intrinsic scatter of at least 0.15 dex is needed to reproduce the \boss clustering which means that \galacticus\ in general shows an insufficient level of scatter. Furthermore, \citet{Tinker17} reported a slightly larger observed scatter of $\sigM=0.18^{+0.01}_{-0.02}$ dex for \cmass and \citet{Leauthaud12_HOD} of $0.249 \pm 0.019$ dex measured from passive galaxies in the \textsc{COSMOS} survey \citep{Scoville07_COSMO}. \citet{Gu16} found similar values for the intrinsic scatter $\sigM<0.2$ and emphasise that the origin of the scatter in the \SHMF\ at higher masses is induced by the hierarchical assembly, while at low halo masses it is associated with in-situ growth. Smaller scatter could mean that there is insufficient scatter in the assembly histories, or that the galaxy formation models do not capture all of it. However, understanding this issue is a non-trivial task and one has to address model specific properties in more detail to understand which combination of properties causes this effect. We find that the comparison with other SAMs would help on this task, but would be beyond the scope of this paper and is therefore left for further studies.

\section{Summary}\label{sec:summary}
Our work is based on the \texttt{Baryon Oscillation Spectroscopic Survey} \citep[\boss,][]{Schlegel09_BOSS,Dawson13_BOSS} of the Sloan Digital Sky Survey \citep[\texttt{SDSS-III},][]{Eisenstein11_SDSS3} \cmass (for ``constant mass'') sample and a semi-analytical model of galaxy formation (SAM), called \galacticus, as part of the \MDG products \citep{Knebe17_MD}. The \cmass sample was build from the \textsc{SDSS-III}/\boss survey catalogues by applying a complex colour-magnitude selection (see \hyperref[eq:cut_dmesa]{\Eq{eq:cut_dmesa}-\EqO{eq:cut_r-i}}). We use the same selection scheme to extract our modelled galaxy catalogue from \galacticus, called \cm, at $z=0.56$.\par

We provide detail assessment of the SAM via comparing with \boss\ as well as results on the galaxy-halo connection and clustering studies of the 2-point correlation function. For reasons stated in \hyperref[sec:modelling]{\Sec{sec:modelling}}, we construct two additional CMASS-mock samples. The first one is called \den\ and was build by randomly selecting modelled galaxies (or {\it{down-sampling}}) until they fit the observational \SMF of \boss between $0.5<z<0.6$. The second CMASS-mock is called \ma and was generated by applying a high stellar mass cut of $\Mstar>10^{11.24}$ \Msun\ as introduced by \citet{Maraston13}. Here we summarise our major results of our study:

\begin{enumerate}
  \item The \galacticus colour-magnitude selected CMASS-mock sample, \cm, shows a lower number density, fewer blue objects, and is located within a smaller parameter space compared to the observational sample (see \hyperref[fig:dperp_i]{\Fig{fig:dperp_i}}). Its red sequence is intrinsically concentrated, as predicted by \citet{Montero-Dorta16} (see \hyperref[fig:r-i_g-r]{\Fig{fig:r-i_g-r}}). Although the number density of this sample is only 1/3 the density of \boss galaxies, \cm overpredicts red galaxies at $\Mstar \gtrsim 10^{12}$ \Msun\ (see \hyperref[fig:smf]{\Fig{fig:smf}}). Galaxies in \den\ satisfy the CMASS colour selection criteria, but they did not enter the sample selection due to their luminosities being approx. 1.5-2 magnitudes lower in $i$-band (see \textit{middle} panel of \hyperref[fig:dis_fidplots]{\Fig{fig:dis_fidplots}}).
  \vspace{0.1cm}

  \item \galacticus \cm and \ma samples agree very well with the stellar to halo mass relation of \citet{Rodriguez-Torres16} and weak-lensing results from \citet{Shan17}, while \den\ shows similar behaviour as the halo abundance matching model from \citet{Behroozi13c} (see \hyperref[fig:SHMF]{\Fig{fig:SHMF}}). However, all three CMASS-mock samples exhibit an increasing scatter at fixed halo mass from $\sigM \sim 0.05-0.15$ dex depending on halo mass. Compared to other works with $\sigM=0.14$ dex by \citet{Rodriguez-Torres16}, $\sigM=0.18^{+0.01}_{-0.02}$ dex by \citet{Tinker17} or $0.249 \pm 0.019$ dex by \citet{Leauthaud12_HOD}, \galacticus\ displays an insufficient level of scatter.
  \vspace{0.1cm}

  \item \cm and \ma agree poorly with the clustering of CMASS galaxies from the high-fidelity mock \blc \citep{Rodriguez-Torres16}, which was obtained using halo abundance matching techniques. We find that the combination of low intrinsic scatter at fixed halo mass and missing objects (or objects being too faint) is responsible for the high clustering amplitudes of \cm and \ma. However, the \denS\ reproduces the clustering of central and satellite galaxies as well as of centrals only, within 1$\sigma$ (see \hyperref[fig:wp]{\Fig{fig:wp}}).
  \vspace{0.1cm}

  \item We can divide the \cm and \den\ samples into two sub-populations, (A) and (B), using a given \SFR\ cut. \textit{Population 
  ~(A)} corresponds to low-starforming galaxies in lower-mass haloes, while \textit{Population~(B)} is comprised by mildly-starforming galaxies living in the most massive haloes. (A)-galaxies were found as the population which displays too faint luminosities as mentioned in (i), but fix the clustering amplitude due to the environmental affiniation and number density. By using the {\sc Vweb} code (see \hyperref[app:A]{\App{app:A}}) we confirm that (A)-galaxies live in filaments, while (B)-galaxies can be found in knots.
  \vspace{0.1cm}

  \item We find further correlations between halo mass \Mc\ and star formation related properties as (specific) star formation rate, gas-phase metallicity, $Z_{\rm cold}$, and cold-gas fraction, $\Mcold/\Mstar$, but also black hole mass $M_{\rm BH}$, depending on the environment and sub-population (A) and (B) where e.g. 80\% of galaxies in \textit{Population~(A)} show higher \sSFR\ and $Z_{\rm cold}>9.5$, but lower cold-gas fractions and black hole masses compared to their counterparts in \textit{Population~(B)} (see \hyperref[tab:env_props]{\Tab{tab:env_props}}). 
\end{enumerate}

\noindent In this work, we have carefully examined several samples of the most massive galaxies from the \galacticus galaxy formation model. In a follow-up work, we plan to extend this analysis to other semi-analytical models in order to study in more detail the star formation history of massive galaxies at intermediate redshifts. This follow-up study will be connected to the effect of \textit{galaxy assembly bias}, a crucial aspect to 
the formation and evolution of galaxies.

\section*{ACKNOWLEDGMENTS}
\noindent DS, FP, ADMD, SRT, GF and AAK want to thank the support of the Spanish Ministry grant AYA2014-60641-C2-1-P managed by the Instituto de Astrof\'isica de Andaluc\'ia (IAA-CSIC).

\noindent WC and AK are supported by the {\it Ministerio de Econom\'ia y Competitividad} and the {\it Fondo Europeo de Desarrollo Regional} (MINECO/FEDER, UE) in Spain through grant AYA2015-63810-P.

\noindent WC further acknowledges the supported by the European Research Council under grant number 670193.

\noindent AK is also supported by the Spanish Red Consolider MultiDark FPA2017-90566-REDC. He further thanks Lance Jyo for dreamwalking.

\noindent ADMD thanks FAPESP for financial support.

\noindent DS fellowship is funded by the \textit{Spanish Ministry of Economy and Competitiveness} (MINECO) under the 2014 \textit{Severo Ochoa} Predoctoral Training Programme. The author also wants to thank the \textit{Bocon\'o Specialty Coffee}-team for their kind supply of energy.

\noindent This work was created by making use by the following software tool and collaborative online platforms: \textsc{Overleaf}\footnote{\url{www.overleaf.com}}, \textsc{matplotlib}\footnote{\url{http://matplotlib.org/}\label{ftn:matplotlib}} 2012-2016, \citet{Matplotlib}; \textsc{Python Software Foundation}\footnote{\url{http://www.python.org}\label{ftn:python}} 1990-2017, version 2.7., \textsc{Pythonbrew}\footnote{\url{https://github.com/utahta/pythonbrew}\label{ftn:pythonbrew}}; \textsc{Cosmolopy}\footnote{\url{http://roban.github.io/CosmoloPy/docAPI/cosmolopy-module.html}\label{ftn:cosmolopy}}; we use whenever possible in this work a colour-blind friendly colour palette\footnote{\url{https://personal.sron.nl/~pault/}\label{ftn:cb-friendly}} for our figures.

\bibliographystyle{mnras}
\bibliography{archive}

\appendix
\section{The Vweb method} \label{app:A}
This Vweb method applies the shear tensor technique
classify the large-scale environments into either ``void'', ``sheet'', ``filament'', or ``knot''.  Following \citet{Hoffman2012}, the velocity shear tensor is defined as
\begin{equation}
 \Sigma_{\alpha\beta} = - \frac{1}{2H_0} \left( \frac{\partial v_\alpha}{\partial r_\beta} + \frac{\partial v_\beta}{\partial r_\alpha} \right),
\end{equation}
where, $H_0$ is the Hubble constant. The eigenvalues of $\Sigma_{\alpha\beta}$ are denoted as $\lambda_i$ ($i$ = 1, 2 and 3).

The simulation box is separated into cubic mesh cells, within which the velocity field is calculated. Each cell has a size of $\sim 1$ Mpc. After smoothing the velocity field, we calculate the eigenvalues of the velocity shear tensor in each cell. Each cell is then classified as either `void', `sheet', `filament', or `knot' according to the eigenvalues $\lambda_1 > \lambda_2 > \lambda_3$:
\begin{itemize}
 \item[1.] void, if $\lambda_1 < \lambda_{th}$,
 \item[2.] sheet, if $\lambda_1 \geq \lambda_{th} > \lambda_2$,
 \item[3.] filament, if $\lambda_2 \geq \lambda_{th} > \lambda_3$,
 \item[4.] knot, if $\lambda_3 \geq \lambda_{th}$,
\end{itemize}
where $\lambda_{th}$ is a free threshold parameter \citep{Hoffman2012, Libeskind2012, Libeskind2013}. Following the discussion of \cite{Carlesi2014, Cui18a, Cui19}, we set $\lambda_{th} = 0.1$, which presents better agreement to the visualised density field. Our mock galaxies are then placed onto the same grid checking for the web classification of the cell they lie in.

\bsp

\label{lastpage}

\end{document}